\documentclass[journal,10pt]{IEEEtran}
\usepackage{cite}
\usepackage{graphicx,color,epsfig,rotating}
\usepackage{amsfonts,amsmath,amssymb,bbm}
\usepackage{algorithm}
\usepackage{algpseudocode}
\usepackage{subfigure}
\usepackage{amsmath}
\usepackage{cite}
\usepackage{placeins}
\usepackage{graphicx}
\usepackage[latin1]{inputenc}
\usepackage{amssymb}
\usepackage{multirow}
\usepackage{stfloats}
\usepackage{tabularx} 
\usepackage{booktabs} 
\usepackage{url}
\usepackage{bm}
\usepackage{soul}
\usepackage{float}
\usepackage{lipsum}     
\usepackage{cuted} 
\usepackage{pstricks}
\usepackage{arydshln}
\usepackage{xspace}

\usepackage{arydshln}
\def\proof{\noindent\hspace{2em}{\it Proof: }}

\usepackage{graphicx}
\usepackage{color}
\usepackage[dvipsnames]{xcolor}

\definecolor{armygreen}{rgb}{0.29, 0.33, 0.13}

\usepackage{amssymb}
\usepackage{amsmath,amsfonts,amssymb}
\usepackage{verbatim}
\usepackage{stfloats}
\usepackage[bookmarks=false]{}
\usepackage{algorithm}
\usepackage{algcompatible}
\usepackage{tikz,pgfplots,tikzpeople}

\newtheorem{theorem}{Theorem}
\newtheorem{corollary}{Corollary}
\newtheorem{definition}{Definition}
\newtheorem{lemma}{Lemma}

\newtheorem{example}{Example}

\begin{document}
\date{}



\title{
Optimal Key Rates for Decentralized Secure Aggregation with Arbitrary Collusion and Heterogeneous Security Constraints
}

\author{
Zhou Li,~\IEEEmembership{Member,~IEEE},
Xiang~Zhang,~\IEEEmembership{Member,~IEEE},
and Giuseppe Caire,~\IEEEmembership{Fellow,~IEEE}

\thanks{Z. Li is with the School of Computer, Electronics and Information, 
Guangxi University, Nanning 530004, China (e-mail: lizhou@gxu.edu.cn).}

\thanks{X. Zhang, and G. Caire are with the Department of Electrical Engineering and Computer Science, Technical University of Berlin, 10623 Berlin, Germany (e-mail: \{xiang.zhang, caire\}@tu-berlin.de).
}




}

\maketitle

\maketitle

\begin{abstract}
Decentralized secure aggregation (DSA) considers a fully-connected network of 
$K$ users, where each pair of users can communicate bidirectionally over an error-free channel. Each user holds a private input, and the goal is for each user to compute the sum of all inputs without revealing any additional information, even in the presence of collusion among up to $T$ users. Traditional DSA typically requires large key sizes to protect all information except for the input sum and the information of colluding users. To mitigate the source keys overhead, we study decentralized secure aggregation with arbitrary collusion and heterogeneous security constraints. In this setting, the inputs of a predefined collection of user subsets, called the \emph{security set} $\bm{\mathcal{S}}$, must be protected from another predefined collection, the \emph{collusion set} $\bm{\mathcal{T}}$. For an arbitrary security set $\mathcal{S}\in \bm{\mathcal{S}}$ and an arbitrary collusion set $\mathcal{T}\in \bm{\mathcal{T}}$, we characterize the optimal communication and source key rates. A key contribution of this work is the characterization of the optimal source key rate, i.e., the minimum number of key bits per input bit that must be shared among users for decentralized secure aggregation with arbitrary collusion and heterogeneous security constraints to be feasible. In general, this characterization reduces to solving a linear program.
\end{abstract}


\allowdisplaybreaks
\section{Introduction}
Federated learning (FL) enables distributed users to collaboratively train a global model without sharing raw data~\cite{mcmahan2017communication,konecny2016federated,kairouz2021advances,yang2018applied}. In each round, users send local updates to a server for aggregation, reducing direct data exposure but still requiring privacy-preserving mechanisms. Decentralized FL (DFL)~\cite{he2018cola,beltran2023decentralized} removes the central server, letting users exchange messages with neighbors. This improves robustness, reduces server dependence, and scales better, motivating primitives like decentralized secure aggregation (DSA) for computing global statistics while preserving input privacy.


Despite not sharing raw data, federated learning (FL) remains vulnerable to privacy attacks such as model inversion~\cite{bouacida2021vulnerabilities, geiping2020inverting, mothukuri2021survey}, where adversaries can reconstruct sensitive information from model updates. This motivates secure aggregation (SA), which ensures that only the sum of user inputs is revealed. Cryptographic SA protocols, beginning with Bonawitz \textit{et al.}~\cite{bonawitz2017practical, bonawitz2016practical} and later improvements~\cite{so2021turbo, kadhe2020fastsecagg, elkordy2022heterosag, jahani2023swiftagg+}, achieve this by generating pairwise masking keys that cancel out at the server. However, these schemes depend on computational assumptions and pseudo-randomness, providing only computational-not information-theoretic-security.

Information-theoretic secure aggregation (IT-SA)~\cite{9834981,zhao2023secure} investigates the fundamental communication and randomness limits of aggregation protocols under perfect security, without relying on computational assumptions. In the classical centralized setting, Zhao and Sun~\cite{zhao2023secure} showed that an optimal scheme achieves communication rate $R_X = 1$, per-user key rate $R_Z = 1$, and total key consumption $R_{Z_\Sigma} = K - 1$, establishing a benchmark for unconditional-security SA.
Subsequent studies extended IT-SA to settings involving user dropout~\cite{so2022lightsecagg}, collusion~\cite{jahani2022swiftagg,jahani2023swiftagg+}, user selection~\cite{zhao2022mds,zhao2023optimal}, groupwise keys~\cite{zhao2023secure,wan2024information,wan2024capacity,Li_Zhang_GroupwiseDSA}, oblivious servers~\cite{sun2023secure}, and hierarchical secure aggregation~\cite{zhang2024optimal,10806947,egger2024privateaggregationhierarchicalwireless,zhang2025fundamental,11195652,li2025collusionresilienthierarchicalsecureaggregation,egger2023private,lu2024capacity}.
Building on these foundations, decentralized secure aggregation (DSA)~\cite{Zhang_Li_Wan_DSA} examines how network topology affects secure aggregation. Unlike centralized schemes, DSA requires all users in a general network graph to recover the global sum simultaneously, introducing new trade-offs between connectivity, communication cost, and key distribution.

Building on DSA, recent work has examined heterogeneous secure aggregation, where only a subset of inputs needs protection~\cite{li2023weakly, li2023weaklysecuresummationcolluding}. This weaker requirement helps reduce randomness consumption. Notably, Zhao and Sun~\cite{zhao2023secure} showed that in classical SA, the optimal communication and key costs remain unchanged even with arbitrary user-server collusion. In contrast, decentralized secure aggregation under heterogeneous security constraint makes the dependence on both the security sets and collusion sets explicit, enabling a finer characterization of how heterogeneous privacy requirements impact the required randomness.

A key challenge in decentralized secure aggregation with
arbitrary collusion and heterogeneous security constraints is
that both the security sets $\bm{\mathcal{S}}$ and collusion sets
$\bm{\mathcal{T}}$ are fully arbitrary, without any layered or nested
structure. Here, each security set $S_m \in \bm{\mathcal{S}}$
specifies users whose inputs must be protected, while each
collusion set $T_n \in \bm{\mathcal{T}}$ represents users that may
collude. Different $S_m$ and $T_n$ induce complex combinatorial
interactions, with binding constraints arising from the most
adversarial $\bm{\mathcal{S}}$--$\bm{\mathcal{T}}$ pairs. Moreover, since
many users lie outside all security sets, key requirements are
decoupled from the input size, with some users needing no keys and
others contributing fractional keys across multiple
constraints, making this setting more challenging than
classical DSA.

Motivated by the rise of decentralized federated learning and the need to mitigate the source keys overhead, we study decentralized secure aggregation with heterogeneous security constraints and arbitrary collusion. Under the honest-but-curious model, we formulate the problem on a fully connected decentralized network and aim to minimize communication and randomness. We characterize the optimal rate region, including the minimum per-link communication and total key cost, showing how the combinatorial structure of the security and collusion sets governs performance and when fractional key assignments are inevitable.



The main contributions of this work are summarized as follows:

\begin{itemize}
\item 
We characterize the exact minimum randomness required for decentralized secure aggregation with arbitrary collusion and heterogeneous security constraints. The optimal key size contains an \emph{integral part} determined by the number of protected users and a \emph{fractional part} given by a linear program.

\item 
We develop a tight converse by converting all security constraints into linear inequalities on key assignments. Identifying the extremal security-collusion pairs yields lower bounds whose fractional component is precisely the LP optimum.

\item 
Guided by the converse structure, we design matching achievable schemes. Integer-rate cases admit direct constructions, while fractional-rate cases are achieved via an LP-based key-splitting method that attains the optimal randomness cost.

\end{itemize}

\section{Problem Statement and Definitions}\label{sec:model}
Consider $K\geq 3$ users, where each user can broadcast its message to all other $K-1$ users without error as shown in Fig.~\ref{fig:model}.
Each user $k\in \{1,2,\cdots,K\}\triangleq [K]$ holds an input vector $W_k$ and a key variable $Z_k$. 
The inputs $W_{1:K}\triangleq\left\{W_k\right\}_{k\in[K]}$ are independent. Each $W_k$ is an $L \times 1$ column vector and the $L$ elements are i.i.d. uniform symbols from the finite field $\mathbb{F}_q$. $W_{1:K}$ is independent of $Z_{1:K}\triangleq\left\{Z_k\right\}_{k\in[K]}$. These conditions are expressed by the equalities:
\begin{eqnarray}
   && H\left(W_{1:K},
    Z_{1:K}\right)=
    \sum_{k\in[K]} H\left(W_k \right) +
    H\left( Z_{1:K} \right), \label{ind} \\
   && H(W_k) = L ~(\mbox{in $q$-ary units}), ~\forall k \in [K]. \label{h2}
\end{eqnarray}
\begin{figure}[t]
    \centering
    \includegraphics[width=0.42\textwidth]{fig_model4.png}
    \vspace{-.2cm}
    \caption{\small Decentralized secure aggregation with five users: key assignment and encoding scheme for Example \ref{ex1}}  
    \label{fig:model}
    \vspace{-.1cm}
\end{figure}
Moreover, the individual keys $Z_{1:K}$ can be arbitrarily correlated \footnote{The use of an arbitrarily correlated key implies an offline key distribution server that prepares and sends the correlated randomness to the users. This step introduces communication overhead, but it is performed only once and does not affect the capacity characterization. This assumption is standard and widely adopted in information theoretic secure computation and secure aggregation.}, 
which necessitates the use of a trusted third-party entity that generates  and distributes the individual keys to each user. 
We assume  that the individual keys are generated from  a \emph{source key} variable $Z_\Sigma$ that contains $L_{Z_\Sigma}$ symbols from $\mathbb{F}_q$ 
\begin{eqnarray}
	H\left(Z_{1:K} \big| Z_\Sigma\right) = 0.\label{total rand}
\end{eqnarray}
To aggregate the inputs, each user $k\in[K]$ generates a message  $X_k$  of length $L_X$ symbols from $\mathbb{F}_q$, using its own input $W_k$ and key $Z_k$. The message $X_k$ is then broadcast to the remaining $K-1$ users. Specifically, we assume $X_k$ is a deterministic function of $W_k$ and $Z_k$,  so that
\begin{eqnarray}
\label{eq:H(Xk|Wk,Zk)=0}
    H(X_k|W_k,Z_k)=0, \forall  k\in[K]. \label{message}
\end{eqnarray}
Each user $u\in[K]$, by combining the messages received from others with its own input and secret key, should be able to recover the sum of all inputs. This leads to the condition:
\begin{align}
\label{correctness}
& \mathrm{[correctness]}\notag\\
& H\Big(\sum_{k\in [K]}W_k \Big|\{X_k\}_{k\in [K]\setminus \{ u\} }, W_u,Z_u\Big  )=0,\forall  u\in[K].
\end{align}
The security input sets are described by a monotone\footnote{A set system is called monotone when, if a set belongs to the system, then its subset also belongs to the system.} set system $\bm{\mathcal{S}}\triangleq \{\mathcal{S}_1, \cdots, \mathcal{S}_M\}$ and the colluding user sets are described by another monotone set\footnote{Without loss of generality, assume $\cup_m \mathcal{S}_m \neq \emptyset$ (the security constraints are not empty) and $|\mathcal{T}_n|\leq K-2$ as otherwise there is nothing to hide.} system $\bm{\mathcal{T}}\triangleq\{\mathcal{T}_1,\cdots, \mathcal{T}_N\}$. The security constraint states that if any user colludes with users from any $\mathcal{T}_n$ set, nothing is revealed about the inputs from any $\mathcal{S}_m$ set (excluding what is possibly already known to users from $\mathcal{T}_n$ when $\mathcal{S}_m \cap \mathcal{T}_n \neq \emptyset$),
\begin{align}
&\mbox{[Security]}~I\Big(\left\{W_k\right\}_{k\in\mathcal{S}_m}; \left\{X_k\right\}_{k\in[K]\setminus \{u\}} \Big| \sum_{k\in [K]} W_k,W_u,\notag\\
& Z_u,\left. \left\{ W_k, Z_k \right\}_{k\in\mathcal{T}_n} \right) = 0, ~\forall m \in [M], n \in [N], u\in [K].\label{security}
\end{align}
In this work, we study both communication and source key rates.
The communication rate $R_X$ characterizes how many symbols each message contains per input symbol, and is defined as
\begin{eqnarray}
    R_X=\frac{L_X}{L}.
\end{eqnarray}
The source key rate $R_{Z_\Sigma}$ characterizes how many symbols the total key variables contains per input symbol, and is defined as
\begin{eqnarray}
    R_{Z_\Sigma} \triangleq \frac{L_{Z_\Sigma}}{L}.
\end{eqnarray}
The rate tuple $(R_X,R_{Z_\Sigma})$ is said to be achievable if there is a decentralized secure aggregation under heterogeneous security Constraints scheme, for which the correctness constraint (\ref{correctness}) and the security constraint (\ref{security}) are satisfied, and the conmmunication rate, and the source key rate are no greater than $R_X$ and $R_{Z_{\Sigma}}$, respectively. The closure of the set of all achievable rate tuples is called the optimal rate region (i.e., capacity region), denoted as $\mathcal{R}^*$.

\subsection{Auxiliary Definitions}
In this section, we introduce several useful definitions that will be used in the subsequent results.


\begin{definition}[Implicit Security Input Set $\mathcal{S}_I$] \label{def:imp} The implicit security input set is defined as 
\begin{align}
\mathcal{S}_I\triangleq& \Big\{ [K]\setminus\{\mathcal{S}_m\cup\mathcal{T}_n\cup\{u\}\} : |\mathcal{S}_m\cup\mathcal{T}_n\cup\{u\}|=K-1,\notag\\
&\forall m \in [M], \forall n \in [N], u\in [K] \Big\} \setminus \{\cup_{i\in [M]}\mathcal{S}_i \}.
\end{align}
\end{definition}
Intuitively, every user in $\cup_{i\in[M]}\mathcal{S}_i$ must hold at least $L$ key symbols to prevent leakage of $W_k$. The implicit security set $\mathcal{S}_I$ further requires that any user outside $\cup_{i\in[M]}\mathcal{S}_i$ but in $\mathcal{S}_I$ must also hold at least $L$ symbols to protect $W_k$. See Corollary~\ref{corollary1} for details. We next use an example to illustrate the definitions.
\begin{example}\label{ex1}
Consider $K=5$, the security input sets are $(\mathcal{S}_1,\mathcal{S}_2,\mathcal{S}_3)=(\emptyset, \{1\}, \{2\})$, and the colluding user sets are $(\mathcal{T}_1,\cdots,\mathcal{T}_{7})=(\emptyset, \{1\}, \{2\}, \{3\}, \{4\},\{5\}, \{2,5\})$.
\end{example}

By searching all $\mathcal{S}_m$, $\mathcal{T}_n$ and user $u$ such that
$|\mathcal{S}_m\cup\mathcal{T}_n\cup\{u\}|=K-1=4$, we find
$|\mathcal{S}_2\cup\mathcal{T}_7\cup\{3\}|=|\{1\}\cup\{2,3,5\}|=4$ and
$|\mathcal{S}_2\cup\mathcal{T}_7\cup\{4\}|=|\{1\}\cup\{2,4,5\}|=4$.
Hence $\mathcal{S}_I=\{3,4\}$ for Example~\ref{ex1}.

\begin{definition}[Total Security Input Set $\overline{\mathcal{S}}$] \label{def:tot} 
The union of explicit and implicit security input sets is defined as the total security input set, 
\begin{eqnarray}
\overline{\mathcal{S}}\triangleq \cup_{m\in[M]} \mathcal{S}_m 
\cup\mathcal{S}_I.
\end{eqnarray}
\end{definition}

Intuitively, every user in $\overline{\mathcal{S}}$ must hold at least $L$ key symbols to protect $W_k$. Since the total security input set $\overline{\mathcal{S}}$ is the union of all explicit and implicit security input sets, any user appearing in either set is therefore required to possess at least $L$ key symbols.
For Example \ref{ex1}, we have $\overline{\mathcal{S}}=\{1\} \cup \{2\} \cup\{3,4\}=\{1,2,3,4\}$.

\begin{definition}[Intersection of $\mathcal{S}_m \cup \mathcal{T}_n \cup \{u\}$ and $\overline{\mathcal{S}}$] \label{def:totset}
Given a security input set $\mathcal{S}_m$, a colluding user set $\mathcal{T}_n$, and a user $u$,
their overlap with the overall security input set $\overline{\mathcal{S}}$ is denoted as
\begin{eqnarray}
\mathcal{A}_{m,n,u}\triangleq (\mathcal{S}_m\cup\mathcal{T}_n\cup\{u\})\cap\overline{\mathcal{S}} 
\end{eqnarray}
and its maximum cardinality is denoted as 
\begin{eqnarray}
a^*\triangleq \max_{m\in[M], n \in [N],u\in [K]} |\mathcal{A}_{m,n,u}|. 
\end{eqnarray}
\end{definition}

\noindent\textit{Intuitively}, users in $\mathcal{A}_{m,n,u}$ are those whose keys jointly appear when masking $W_k$. Each must contribute $L$ symbols and their keys cannot overlap, so the group requires $|\mathcal{A}_{m,n,u}|L$ distinct symbols. Hence $a^*$ reflects the worst-case size of such a group.


For Example \ref{ex1}, $\mathcal{A}_{2,7,3}= (\{1\}\cup\{2,5\}\cup\{3\})\cap\{1,2,3,4\} = \{1,2,3\}$, $\mathcal{A}_{2,7,4}= (\{1\}\cup\{2,5\}\cup\{4\})\cap\{1,2,4\} = \{1,2,4\}$, $\mathcal{A}_{3,2,3}= (\{2\}\cup\{1\}\cup\{3\})\cap\{1,2,3,4\} = \{1,2,3\}$, and $\mathcal{A}_{3,2,5}= (\{2\}\cup\{1\}\cup\{5\})\cap\{1,2,3,4\} = \{1,2\}$. Hence, $a^*=3$.

\begin{definition}[Union of Maximum $\mathcal{A}_{m,n,u}$] \label{def:uni}
Find all $\mathcal{A}_{m,n,u}$ sets with maximum cardinality
and denote the union of the corresponding $\mathcal{S}_m, \mathcal{T}_n, \{u\}$ sets as $\mathcal{Q}$.
\begin{eqnarray}
\mathcal{Q} \triangleq \cup_{m,n,u: |\mathcal{A}_{m,n,u}| = a^*} \mathcal{S}_m \cup \mathcal{T}_n\cup\{u\}. 
\end{eqnarray}
\end{definition}
For Example \ref{ex1}, $\mathcal{A}_{2,3,3}, \mathcal{A}_{2,3,4},\mathcal{A}_{3,2,3},\mathcal{A}_{3,2,4},\mathcal{A}_{2,7,3},\mathcal{A}_{2,7,4}$ are all $\mathcal{A}_{m,n,u}$ sets with the maximum cardinality, so 
$\mathcal{Q} = (\mathcal{S}_2 \cup \mathcal{T}_{3}\cup\{3\}) \cup (\mathcal{S}_2 \cup \mathcal{T}_{3}\cup\{4\}) \cup (\mathcal{S}_3 \cup \mathcal{T}_{2}\cup\{3\}) \cup (\mathcal{S}_3 \cup \mathcal{T}_{2}\cup\{4\}) \cup (\mathcal{S}_2 \cup \mathcal{T}_{7}\cup\{3\}) \cup (\mathcal{S}_2 \cup \mathcal{T}_{7}\cup\{4\}) 
= \{1,2,3,4\}$.

\section{Result}
\begin{theorem}\label{thm:d1}
For decentralized secure aggregation with $K \geq 3$ users, security input sets $\{\mathcal{S}_m\}_{m\in[M]}$, and colluding user sets $\{\mathcal{T}_n\}_{n\in[N]}$, 
the optimal rate region $\mathcal{R}^*$ is 
\begin{align}
   & \mathcal{R}^* =\{ \left(R_X, R_{Z_\Sigma}\right) |  
R_X\ge 1,  R_{Z_{\Sigma}} \geq R^*_{Z_{\Sigma}}\} \notag
\end{align}
where $R^*_{Z_{\Sigma}}$ is given by
\begin{align}
   & R^*_{Z_{\Sigma}} =
\begin{cases}
K - 1 , &  \text{if}~ a^* = K,\\
~~~a^* ,
&\text{else if}~  a^* < \big|\overline{\mathcal{S}}\big|, ~\text{or} \\
&  a^* = \big|\overline{\mathcal{S}}\big| \text{and}~ |\mathcal{Q}| < K\\
a^* + b^*, & \text{otherwise } 
\end{cases}
\end{align}
where $b^*$ is the optimal value of the following linear program\footnote{Note that for the `If' case, $|\mathcal{S}_m \cup \mathcal{T}_n\cup \{u\}| \leq K-1$ because otherwise $a^* = K$; for all $m,n,u$ such that $|\mathcal{A}_{m,n,u}| = a^*$, $a^* = |\overline{\mathcal{S}}|$ so $\overline{\mathcal{S}} \subset (\mathcal{S}_m \cup \mathcal{T}_n\cup \{u\})$ and $b_k, k \in [K] \setminus \overline{\mathcal{S}}$ are all the variables.}, 
\end{theorem}

\begin{align}
&\min \max_{m,n,u: |\mathcal{A}_{m,n,u}| = a^* }  \sum_{k\in (\mathcal{T}_{n}\cup\{u\})\setminus\overline{\mathcal{S}}} b_k \notag
\\
 &\text{subject to} \notag\\
    & \sum_{k\in [K]\setminus (\mathcal{S}_m \cup \mathcal{T}_n\cup\{u\})} b_k 
    \geq 1,\forall m,n,u ~\mbox{such that}~ |\mathcal{A}_{m,n,u}| = a^*, 
    \notag \\
    & b_k\geq 0, \forall k\in [K]\setminus\overline{\mathcal{S}}.
    \label{lp}
\end{align}

For all cases, the minimum communication rate is $1$. Intuitively, the communication cost is dictated by the input $W_k$, because to satisfy the correctness constraint~(\ref{correctness}), each user $k$ must send at least $L$ symbols to every other user $u \in [K]\setminus\{k\}$, such that the message received by user $u$ contains the information of $W_k$ to get the sum $\sum_{k\in [K]}W_k$

We next analyze the source key rate for each case.

\textbf{`If' case ($a^* = K$):}
In this case, $|\mathcal{S}_m \cup \mathcal{T}_n \cup \{u\}| = |\overline{\mathcal{S}}| = K$, which ensures that every user must hold at least $L$ key symbols. Hence $R_{Z_\Sigma} \ge K-1,$
and the required source key rate coincides with that of the classical decentralized secure aggregation scheme in~\cite{Zhang_Li_Wan_DSA}.

\textbf{`Else if' case ($a^* \le K-1$):}
This case is divided into two subcases.
\begin{itemize}
    \item \textbf{Subcase 1 ($a^* < |\overline{\mathcal{S}}|$):}
    $R_{Z_\Sigma} \ge a^*.$
    Since $a^*$ is strictly smaller than $|\overline{\mathcal{S}}|$, it suffices to assign keys only to the users in $\overline{\mathcal{S}}$, resulting in a total key size of $a^* L$.

    \item \textbf{Subcase 2 ($a^* = |\overline{\mathcal{S}}|$, and $|\mathcal{Q}| \le K-1$):}
    $R_{Z_\Sigma} \ge a^*.$
    Each user in $\overline{\mathcal{S}}$ receives $L$ key symbols. One additional user in $[K]\setminus \mathcal{Q}$ also receives $L$ key symbols. These additional key symbols are linear combinations of those assigned to the users in $\overline{\mathcal{S}}$, and their sum is zero to satisfy correctness. Therefore, the number of independent key symbols remains $a^* L$.
\end{itemize}

\textbf{`Otherwise' case ($a^* \le K-1$, $a^* = |\overline{\mathcal{S}}|$, and $|\mathcal{Q}| = K$):}
In this case,
$R_{Z_\Sigma} \ge a^* + b^*.$
Each user in $\overline{\mathcal{S}}$ is assigned $L$ key symbols, and each user in $[K]\setminus \overline{\mathcal{S}}$ receives $b_k^* L$ key symbols with $b_k^* \le 1$. In particular,
$a^* + b^*
= |\overline{\mathcal{S}}|
+ \sum_{k \in (\mathcal{T}_n \cup \{u\}) \setminus \overline{\mathcal{S}}} b_k^*
\le |\overline{\mathcal{S}}|
+ |(\mathcal{T}_n \cup \{u\}) \setminus \overline{\mathcal{S}}|
\le K-1.$

In summary, for the `If'and ``else if'' cases, whenever a user is assigned the key values, its key size must be $L$ symbols, and at most one user in $[K]\setminus \mathcal{Q}$ receives $L$ symbols. All key sizes in these two cases are integers. In contrast, in the ``otherwise'' case, each user in $[K]\setminus \mathcal{Q}$ is assigned only $b_k^* L$ symbols, which can be strictly smaller than $L$. Although the total source key cost depends on the predefined security set $\bm{\mathcal{S}}$ and collusion set $\bm{\mathcal{T}}$, it is always no larger than that of the classical DSA scheme, namely $K-1$. Moreover, in the `If' case, the optimal $b_k^*$ values are determined by the linear program~(\ref{lp}), and the scheme allocates exactly $b_k^* L$ key symbols to each user, thereby achieving the converse bound.

\section{Converse Proof of Theorem \ref{thm:d1}} \label{sec:converse}
For the converse proof, the parameter $a^*$ plays a crucial role. Specifically, we first establish that $R_{Z_\Sigma} \geq a^*$ when $a^*\leq K-1$, and $R_{Z_\Sigma} \geq K-1$ when $a^*=K$. For all cases, $R_X\geq 1$.
To illustrate this idea, we refer to Example \ref{ex1}

\subsection{Converse Proof of Example \ref{ex1}} \label{sec:ex1}
Consider the Example \ref{ex1}, $a^*=3\leq K-1=4$, we show that $R_{Z_\Sigma} \geq a^*=3$ and $R_X\geq 1$.

We first present an auxiliary result showing that the entropy of $X_1$, conditioned on the other users' inputs and keys, is no less than $L$ symbols:
\begin{eqnarray}
\label{eq:result1,H(X1|(Wi,Zi)i=2,3)>=L,example}
H(X_1)\geq H\left(X_1 \mid \{W_k,Z_k\}_{k\in \{2,3,4,5\}}\right) \geq L.
\end{eqnarray}
The intuition is as follows. Each user $k\in \{2,3,4,5\}$ is required to recover the aggregate of all inputs except its own, which necessarily includes $W_1$. Since $W_1$ is available exclusively at User 1, the information must be embedded in $X_1$. Consequently, the entropy of $X_1$ cannot be smaller than that of $W_1$, namely $L$.
The formal argument proceeds as
\begin{align}
H(X_1)\geq& H\left(X_1| \{W_k,Z_k\}_{k\in \{2,3,4,5\}}  \right) \notag \\ 
  \ge & I\left(X_1; W_1|\{W_k,Z_k\}_{k \in\{2,3,4,5\} }\right)\\
 = & H\left(W_1|\{W_k,Z_k\}_{k \in\{2,3,4,5\} }\right) \notag\\
 &- H\left(W_1|\{W_k,Z_k\}_{k \in\{2,3,4,5\}},X_1\right) \\
 \overset{(\ref{ind})}{=} & H\left(W_1\right) - H\left(W_1|\{W_k,Z_k\}_{k \in\{2,3,4,5\}},\right.\notag\\
 &\left. X_1,W_1+W_2+W_3+W_4+W_5\right)
\label{eq:step 0, proof of lemma, example}\\
 = &  L.\label{eq:step 1, proof of lemma, example}
\end{align}
In (\ref{eq:step 0, proof of lemma, example}), the first term comes from the independence between inputs and keys. The second follows because $X_2,X_3,X_4,X_5$ are deterministic functions of their respective $\{W_k,Z_k\}_{k\in \{2,3,4,5\}}$, and the overall sum $W_1+W_2+W_3+W_4+W_5$ can be reconstructed from $(X_1,\dots,X_5)$. Finally, (\ref{eq:step 1, proof of lemma, example}) holds since $W_1$ is uniquely determined by $(W_2,W_3,W_4,W_5)$ together with the global sum $W_1+W_2+W_3+W_4+W_5$.

Symmetrically, we can also obtain
\begin{eqnarray}
    \label{eq:result1,H(X2|(Wi,Zi)i=1,3)>=L,example}
H(X_u)\geq H\left(X_u|\{W_k,Z_k\}_{k\in [5]\setminus\{u\}}\right) \ge L, \forall u\in[5].
\end{eqnarray}

Consider the  communication rate, we have
\begin{eqnarray}
    R_X\triangleq \frac{L_{X}}{L} \geq \frac{H(X_1)}{L} \overset{(\ref{eq:step 1, proof of lemma, example})}{\geq} 1.
\end{eqnarray}

Next, consider the set $\mathcal{A}_{m,n,u}$. We aim to show that $H(Z_{\Sigma}) \;\geq\; H\big(\{Z_k\}_{k\in \mathcal{A}_{m,n,u}}\big) 
    \;\geq\; |\mathcal{A}_{m,n,u}|L .$
In Example~\ref{ex1}, for instance, let $\mathcal{A}_{1,7,3}=\{1,2,3\}$. 
In this case, it suffices to prove that 
\begin{eqnarray}
    H(Z_1,Z_2,Z_3) \;\geq\; 3L .
\end{eqnarray}
Note that user $3$ belongs to the implicit security input set $\mathcal{S}_I$. 
Therefore, we first show that 
    $H(Z_3) \;\geq\; L ,$
which establishes the desired lower bound.
\begin{align}
&H(Z_3) \geq H\left(Z_{3}\big|\{Z_k\}_{k\in \mathcal{T}_{7}},Z_4\right) = H(Z_3|Z_2,Z_4,Z_5)\notag\\
\geq&I(Z_3;Z_1|Z_2,Z_4,Z_5)\\
\overset{(\ref{ind})}{=}& I(Z_3,W_3;Z_1,W_1|\{Z_k,W_k\}_{k\in \{2,4,5\}}) \label{eq:t1}\\
\overset{(\ref{message})}{\geq}&I(X_3,W_3;X_1,W_1|\{Z_k,W_k\}_{k\in \{2,4,5\}}) \\
\geq& I(X_3,W_3;W_1|\{Z_k,W_k\}_{k\in \{2,4,5\}},X_1) \label{eq:t2}\\
=&H(W_1|\{Z_k,W_k\}_{k\in \{2,4,5\}},X_1)\notag\\
&-H(W_1|\{Z_k,W_k\}_{k\in \{2,4,5\}},X_1,X_3,W_3)\\
\overset{(\ref{message})(\ref{correctness})}{\geq}&H\Big(W_1\Big|\{Z_k,W_k\}_{k\in \{2,4,5\}}, \{X_k\}_{k\in[5]}, \sum_{k\in[5]} W_k\Big) \notag\\
&-~H\Big(W_1\Big|\{Z_k,W_k\}_{k\in \{2,4,5\}},\{X_k\}_{k\in[5]},W_3, \sum_{k\in[5]} W_k\Big) \label{eq:t3}\\
\geq&H\Big(W_1\Big|\{Z_k,W_k\}_{k\in \{2,4,5\}}, \sum_{k\in[5]} W_k\Big)\notag\\
&-\underbrace{I\Big(W_1;\{X_k\}_{k\in[5]\setminus\{4\}}\Big|\{Z_k,W_k\}_{k\in \{2,4,5\}}, \sum_{k\in[5]} W_k\Big)}_{\overset{(\ref{security})}{=}0}\notag\\
&-H(W_1|W_1) \label{eq:t4}\\
\overset{(\ref{ind})(\ref{h2})}{=}&L - 0 = L, \label{impsec1}
\end{align}
where (\ref{eq:t1}) follows from the independence between $\{W_k\}_{k\in[K]}$ and $\{Z_k\}_{k\in[K]}$ (see (\ref{ind})), and (\ref{eq:t2}) follows since $X_k$ is a deterministic function of $(W_k, Z_k)$ (see (\ref{message})). In (\ref{eq:t3}), the first term is obtained by conditioning additionally on all $\{X_k\}_{k\in [K]}$ and the desired sum $\sum_{k\in [K]} W_k$; the second term follows from the correctness constraint (\ref{correctness}), which guarantees that $\sum_{k\in [K]} W_k$ can be decoded from all $\{X_k\}_{k\in [K]}$. In (\ref{eq:t4}), the second term vanishes due to the security constraint (\ref{security}), where the security input set is $\mathcal{S}_2=\{1\}$ and user~4 colludes with the collusion set $\mathcal{T}_7=\{2,5\}$. The third term is obtained by retaining only $W_1$, which can be recovered from $\sum_{k\in [K]} W_k$ together with $(W_2, W_3, W_4, W_5)$. Finally, the last step uses both the independence of $\{W_k\}_{k\in[K]}$ and $\{Z_k\}_{k\in[K]}$ (see (\ref{ind})) and the uniform distribution of $\{W_k\}_{k\in[K]}$ (see (\ref{h2})). Intuitively, even though each user in the implicit set that does not belong to the union of the security input sets $\cup_{k\in[M]}\mathcal{S}_i$ should still hold at least $L$ key symbols.

Second, consider $H(Z_1,Z_2|Z_3) \geq 2L$. Note that $\mathcal{S}_2\cup\mathcal{T}_{7}\cup\{3\}=\{1\}\cup\{2,5\}\cup\{3\}=\{1,2,3,5\}$ so that $Z_3$ may appear in the conditioning term and $(\mathcal{S}_2\cup\mathcal{T}_{7}\cup\{3\})\cap\cup_{m\in [3]}\mathcal{S}_m=\{1,2,3,5\}\cap\{1,2\}=\{1,2\}$ so that we want to show that $Z_1,Z_2$ must each contribute $L$ independent amount of information.
\begin{align}
& H(Z_1,Z_2|Z_3)\notag\\
\geq&H(Z_1,Z_2|Z_3,W_1,W_2,W_3)\\
\geq&I(Z_1,Z_2;X_1,X_2|Z_3,W_1,W_2,W_3)\\
=&H(X_1,X_2|Z_3,W_1,W_2,W_3)\notag\\
&-H(X_1,X_2|Z_3,W_1,W_2,W_3,Z_1,Z_2)\\
\overset{(\ref{message})}{=}&H(X_1,X_2|W_3,Z_3)-I(X_1,X_2;W_1,W_2|W_3,Z_3)\\
=&H(X_1,X_2|W_3,Z_3)- I(X_1,X_2;W_1|W_3,Z_3) \notag\\
&- I(X_1,X_2;W_2|W_3,Z_3,W_1)  \label{sec_indi}\\
\geq&H(X_1|W_3,Z_3)+H(X_2|X_1,Z_1,W_1,W_3,Z_3)\notag\\
&- I\Big(X_1,X_2, \sum_{k\in[5]} W_k ;W_1\Big|W_3, Z_3\Big) \notag\\
&- I\Big(X_1,X_2,\sum_{k\in[5]} W_k, Z_1;W_2\Big|W_3,Z_3,W_1\Big) \\
\overset{(\ref{message})(\ref{ind})}{=}&H(X_1|W_3,Z_3)+H(X_2|Z_1,W_1,W_3,Z_3)\notag\\
&- \underbrace{I\Big(X_1,X_2 ;W_1\Big| \sum_{k\in[5]} W_k, W_3, Z_3\Big)}_{\overset{(\ref{security})}{=}0} \notag\\
&- \underbrace{I\Big(X_1,X_2;W_2\Big|\sum_{k\in[5]} W_k, Z_1, W_3,Z_3,W_1\Big)}_{\overset{(\ref{security})}{=}0} \label{totalsec1} \\
\overset{(\ref{eq:result1,H(X1|(Wi,Zi)i=2,3)>=L,example})(\ref{eq:result1,H(X2|(Wi,Zi)i=1,3)>=L,example})}{\geq}&2L, 
\end{align}
where in (\ref{totalsec1}), the second term follows since $X_1$ is a function of $(W_1,Z_1)$. The third and fourth terms follow from the independence of $\{W_k\}_{k\in[K]}$ and $\{Z_k\}_{k\in[K]}$. In particular, in (\ref{totalsec1}), the third term vanishes due to the security constraint (\ref{security}) for user 3 with $\mathcal{S}_2=\{1\}$ and $\mathcal{T}_0=\emptyset$, while the fourth term vanishes due to the security constraint (\ref{security}) for user 3 with $\mathcal{S}_3=\{2\}$ and $\mathcal{T}_2=\{1\}$.

\subsection{Key Lemmas and Corollaries}
We are now ready to generalize the above proof to all parameter settings. Let's start with a few useful lemmas. 
First as a preparation, we show that each $X_k$ must contain at least $L$ symbols (the size of the input) even if all other inputs are known.
\begin{lemma}\label{lemma1}
For any $u \in [K]$, we have 
\begin{eqnarray}
    H\left(X_u\big|\{W_k,Z_k\}_{k\in[K]\backslash \{u\}}\right) \geq L. \label{lemma1_eq}
\end{eqnarray}
\end{lemma}
Intuitively, each user $k$ needs to recover the aggregate of all inputs excluding its own, which necessarily involves $W_u$ for all $u \in [K]\setminus {k}$. Since $W_u$ is only available at User $u$, the corresponding information must be contained in $X_u$. Therefore, the entropy of $X_u$ cannot be smaller than that of $W_u$, i.e., $H(X_u) \ge L$. Formally, we have

\proof
\begin{align}
& H\left(X_u\big|\{W_k,Z_k\}_{k\in[K]\backslash \{u\}}\right)\notag\\
\geq& I\left(X_u; W_u \big|\{W_k,Z_k\}_{k\in[K]\backslash \{u\}}\right)\\
 =& H\left( W_u \big| \{W_k,Z_k\}_{k\in[K]\backslash \{u\}}\right) \notag\\
 &- H\left(W_u \big| X_u, \{W_k,Z_k\}_{k\in[K]\backslash \{u\}}\right)\\
\overset{(\ref{ind})(\ref{message})}{\geq}& H\left(W_u\right)-\notag\\
& H\Big(W_u \Big| \{X_k\}_{k\in [K]},\sum_{k\in[K]}W_k,\{W_k\}_{k\in[K]\backslash \{u\}}\Big)\label{pf_lemma1_1}\\
\overset{(\ref{h2})}{=}& L. \label{eq:e1} 
\end{align}
In (\ref{pf_lemma1_1}), the first term follows from the independence of $W_u$ from the other inputs and keys $\{W_k,Z_k\}_{k\in[K]\setminus\{u\}}$ (see (\ref{ind})). The second term follows from the fact that $\{X_k\}_{k\in[K]\setminus\{u\}}$ is determined by $\{W_k,Z_k\}_{k\in[K]\setminus\{u\}}$ (see (\ref{message})), and it equals zero because $W_u$ can be recovered from $\sum_{k\in[K]} W_k$ given $\{W_k\}_{k\in[K]\setminus\{u\}}$. In (\ref{eq:e1}), note that $W_u$ consists of $L$ independent and uniformly distributed symbols (see (\ref{h2})).

Next, we show that, due to the correctness constraint, the keys used by users outside $\mathcal{S}_m \cup \mathcal{T}_n \cup \{u\}$ must contain at least $L$ symbols, conditioned on the information known to the colluding users.
\begin{lemma} \label{lemma2} 
For any $\mathcal{S}_m, \mathcal{T}_n$ with $m\in[M], n\in[N]$ and any $u\in[K]$ such that $\mathcal{S}_m \cap (\mathcal{T}_n \cup \{u\}) = \emptyset$ and $|\mathcal{S}_m \cup \mathcal{T}_n \cup \{u\}| \le K-1$, we have
\begin{eqnarray}
    H\left(\{Z_k\}_{k \in [K] \setminus(\mathcal{S}_m\cup\mathcal{T}_n\cup\{u\})}\big|\{Z_k\}_{k\in\mathcal{T}_n\cup\{u\}}\right) \geq L. \label{lemma2_eq}
\end{eqnarray}
\end{lemma}
Intuitively, to satisfy both the security and correctness constraints, the total key size of users in the set $[K] \setminus (\mathcal{S}_m \cup \mathcal{T}_n \cup \{u\})$ must be at least $L$ symbols, even when the keys of users in $\mathcal{T}_n \cup \{u\}$ are known.

\proof 
\begin{align}
&H\left(\{Z_k\}_{k \in [K] \setminus(\mathcal{S}_m\cup\mathcal{T}_n\cup\{u\})}\big|\{Z_k\}_{k\in\mathcal{T}_n\cup\{u\}}\right) \notag\\
\geq&I\left(\{Z_k\}_{k \in [K] \setminus(\mathcal{S}_m\cup\mathcal{T}_n\cup\{u\})};\{Z_k\}_{k \in\mathcal{S}_m}\big|\{Z_k\}_{k\in\mathcal{T}_n\cup\{u\}}\right)\\
\overset{(\ref{ind})}{=}&I\left(\{Z_k,W_k\}_{k \in [K] \setminus(\mathcal{S}_m\cup\mathcal{T}_n\cup\{u\})};\{Z_k,W_k\}_{k \in\mathcal{S}_m}\big|\right.\notag\\
&\left. \{Z_k,W_k\}_{k\in\mathcal{T}_n\cup\{u\}}\right) \label{eq:tx1}\\
\overset{(\ref{message})}{\geq}&I\left(\{X_k,W_k\}_{k \in [K] \setminus(\mathcal{S}_m\cup\mathcal{T}_n\cup\{u\})};\{X_k,W_k\}_{k \in\mathcal{S}_m}\big|\right.\notag\\
&\left. \{Z_k,W_k\}_{k\in\mathcal{T}_n\cup\{u\}}\right)\\
\geq&I\left(\{X_k,W_k\}_{k \in [K] \setminus(\mathcal{S}_m\cup\mathcal{T}_n\cup\{u\})};\{W_k\}_{k \in\mathcal{S}_m}|\right.\notag\\
&\left.\{Z_k,W_k\}_{k\in\mathcal{T}_n\cup\{u\}},\{X_k\}_{k \in\mathcal{S}_m}\right)\\
=&H\left(\{W_k\}_{k \in\mathcal{S}_m}\big|\{Z_k,W_k\}_{k\in\mathcal{T}_n\cup\{u\}},\{X_k\}_{k \in\mathcal{S}_m}\right)\notag \\
&-~H\left(\{W_k\}_{k \in\mathcal{S}_m}\big|\{Z_k,W_k\}_{k\in\mathcal{T}_n\cup\{u\}},\{X_k\}_{k \in\mathcal{S}_m},\right.\notag\\
&\left.\{X_k,W_k\}_{k \in [K] \setminus(\mathcal{S}_m\cup\mathcal{T}_n\cup\{u\})}\right)\\
\overset{(\ref{correctness})}{\geq}&H\Big(\{W_k\}_{k \in\mathcal{S}_m}\Big|\sum_{k\in [K]} W_k, \left( W_k, Z_k \right)_{k\in\mathcal{T}_n\cup\{u\}}, \{X_k\}_{k \in[K]} \Big)\notag\\
&-H\Big(\{W_k\}_{k \in\mathcal{S}_m}\Big|\sum_{k\in \mathcal{S}_m} W_k\Big)\label{eq:lem2t1}\\
\overset{(\ref{security})}{=}&H\Big(\{W_k\}_{k \in\mathcal{S}_m}\Big|\sum_{k\in [K]} W_k, \left( W_k, Z_k \right)_{k\in\mathcal{T}_n\cup\{u\}}\Big)-\notag\\
&\underbrace{I\Big(\{W_k\}_{k \in\mathcal{S}_m}; \{X_k\}_{k \in[K]}\Big|\sum_{k\in [K]} W_k, \left( W_k, Z_k \right)_{k\in\mathcal{T}_n\cup\{u\}} \Big)}_{\overset{(\ref{security})}{=}0}\notag\\
&-H\Big(\{W_k\}_{k \in\mathcal{S}_m}\Big|\sum_{k\in \mathcal{S}_m} W_k\Big)\label{eq:lem2t2}\\
\overset{(\ref{h2})}{=}&|\mathcal{S}_m| L-(|\mathcal{S}_m|-1) L \label{eq:tt2}\\
=&L.
\end{align}
In (\ref{eq:lem2t1}), the first term is obtained by further conditioning on all transmitted messages $\{X_k\}_{k\in[K]}$ and the desired sum $\sum_{k\in[K]} W_k$. The second term follows from the correctness constraint~(\ref{correctness}): given $\{X_k\}_{k\in[K]}$, the sum $\sum_{k\in[K]} W_k$ can be recovered, and the values $\{W_k\}_{k\in\mathcal{S}_m}$ are fully determined by $\sum_{k\in[K]} W_k$, $\{W_k\}_{k\in\mathcal{T}_n\cup\{u\}}$, and $\{W_k\}_{k\in[K]\setminus(\mathcal{S}_m\cup\mathcal{T}_n\cup\{u\})}$. Moreover, the second term in (\ref{eq:lem2t2}) vanishes due to the security constraint~(\ref{security}).

Lemma~\ref{lemma2}, evaluated in the setting where $|\mathcal{S}_m\cup\mathcal{T}_n\cup\{u\}| = K-1$, immediately yields the corollary stated next. This result pertains to the elements of the implicit security input set $\mathcal{S}_{I}$, defined in Definition~\ref{def:imp}.
\begin{corollary} \label{corollary1}
For any $\mathcal{S}_m$, $\mathcal{T}_n$ with $m\in[M]$, $n\in[N]$, and any $u\in[K]$ such that $\mathcal{S}_m \cap (\mathcal{T}_n \cup \{u\}) = \emptyset$ and $|\mathcal{S}_m \cup \mathcal{T}_n \cup \{u\}| = K-1$, let $u' = [K]\setminus(\mathcal{S}_m \cup \mathcal{T}_n \cup \{u\})$, and we have 
\begin{eqnarray}
    H\left(Z_{u'}\big|\{Z_k\}_{k\in\mathcal{T}_n\cup\{u\}}\right) \geq L. \label{corollary1_eq}
\end{eqnarray}
\end{corollary}
Intuitively, each user in $\mathcal{S}_I$ must possess at least $L$ symbols, even if the keys of the users in $\mathcal{T}_n \cup {u}$ are revealed.

For any $\mathcal{S}_m$, $\mathcal{T}_n$ and $u$, we show that, given the information available to the colluding users, the total number of key symbols held by users in $(\mathcal{S}_m\cup\mathcal{T}_n\cup\{u\})\cap(\cup_{k\in[M]}\mathcal{S}_i)$ is at least $L$ times the cardinality of that set.
\begin{lemma} \label{lemma3} 
Let $m\in [M]$, $n\in [N]$, and $u\in [K]$. For any $\mathcal{S}_m$ and $\mathcal{T}_n$ satisfying 
$\mathcal{S}_m \cap (\mathcal{T}_n \cup \{u\}) = \emptyset$ and 
$|\mathcal{S}_m \cup \mathcal{T}_n \cup \{u\}| \le K-1$, we have
\begin{align}
   &H\big(\{Z_k\}_{k \in \mathcal{U}_{m,n,u}\cap \mathcal{S}_{[M]}} \,\big|\,
     \{Z_k\}_{k \in (\mathcal{T}_n \cup \{u\}) \setminus \mathcal{S}_{[M]}}\big) \notag\\
    \geq& |\mathcal{U}_{m,n,u} \cap \mathcal{S}_{[M]}| L, \label{lemma3_eq}
\end{align}
where $\mathcal{S}_{[M]} \triangleq \cup_{k\in [M]} \mathcal{S}_k$ and 
$\mathcal{U}_{m,n,u} \triangleq \mathcal{S}_m \cup \mathcal{T}_n \cup \{u\}$.
\end{lemma}

\proof
\begin{align}
&H\big(\{Z_k\}_{k \in \mathcal{U}_{m,n,u}\cap\mathcal{S}_{[M]}}\big|\{Z_k\}_{k\in(\mathcal{T}_n\cup\{u\})\setminus \mathcal{S}_{[M]}}\big) \notag\\
\geq& I\big(\{Z_k\}_{k \in \mathcal{U}_{m,n,u}\cap\mathcal{S}_{[M]}}; \{X_k\}_{k \in \mathcal{U}_{m,n,u}\cap\mathcal{S}_{[M]}}\big|\notag\\
& \{Z_k\}_{k\in(\mathcal{T}_n\cup\{u\})\setminus \mathcal{S}_{[M]}}, \{W_k\}_{k \in \mathcal{U}_{m,n,u}\cap\mathcal{S}_{[M]}}\big) \\
\overset{(\ref{message})}{=}&H\big(\{X_k\}_{k \in \mathcal{U}_{m,n,u}\cap\mathcal{S}_{[M]}}\big|\notag\\
&\{Z_k\}_{k\in(\mathcal{T}_n\cup\{u\})\setminus \mathcal{S}_{[M]}},\{W_k\}_{k \in \mathcal{U}_{m,n,u}\cap\mathcal{S}_{[M]}}\big) \\
=&H\big(\{X_k\}_{k \in \mathcal{U}_{m,n,u}\cap\mathcal{S}_{[M]}}\big| \{Z_k\}_{k\in(\mathcal{T}_n\cup\{u\})\setminus \mathcal{S}_{[M]}}\big) \notag\\
&-I\big(\{X_k\}_{k \in \mathcal{U}_{m,n,u}\cap\mathcal{S}_{[M]}}; \{W_k\}_{k \in \mathcal{U}_{m,n,u}\cap\mathcal{S}_{[M]}}\big|\notag\\
&\{Z_k\}_{k\in(\mathcal{T}_n\cup\{u\})\setminus \mathcal{S}_{[M]}}\big) \\
\geq& H\big(\{X_k\}_{k \in \mathcal{U}_{m,n,u}\cap\mathcal{S}_{[M]}}\big|\{Z_k\}_{k\in(\mathcal{T}_n\cup\{u\})\setminus \mathcal{S}_{[M]}}\big) \notag\\
&-~I\Big(\{X_k\}_{k \in [K]},\sum_{k\in [K]}W_k;\{W_k\}_{k \in \mathcal{U}_{m,n,u}\cap\mathcal{S}_{[M]}}\big|\notag\\
&\{Z_k\}_{k\in(\mathcal{T}_n\cup\{u\})\setminus \mathcal{S}_{[M]}}\big) \\
\geq& H\big(\{X_k\}_{k \in \mathcal{U}_{m,n,u}\cap\mathcal{S}_{[M]}}\big|\{Z_k\}_{k\in(\mathcal{T}_n\cup\{u\})\setminus \mathcal{S}_{[M]}}\big)- \notag\\
&I\Big(\sum_{k\in [K]}W_k;\{W_k\}_{k \in \mathcal{U}_{m,n,u}\cap\mathcal{S}_{[M]}}\Big|\{Z_k\}_{k\in(\mathcal{T}_n\cup\{u\})\setminus \mathcal{S}_{[M]}}\Big) \notag\\
&-~I\left(\{X_k\}_{k \in [K]};\{W_k\}_{k \in \mathcal{U}_{m,n,u}\cap\mathcal{S}_{[M]}}\big|\right.\notag\\
&\sum_{k\in [K]}W_k,\{Z_k\}_{k\in(\mathcal{T}_n\cup\{u\})\setminus \mathcal{S}_{[M]}}\Big) \\
\geq&H\big(\{X_k\}_{k \in \mathcal{U}_{m,n,u}\cap\mathcal{S}_{[M]}}\big|\{Z_k\}_{k\in(\mathcal{T}_n\cup\{u\})\setminus \mathcal{S}_{[M]}}\big) \notag\\
&-~I\left(\{X_k\}_{k \in [K]}; (W_{k})_{k\in(\mathcal{T}_n\cup\{u\}) \cap \mathcal{S}_{[M]}} \big|\right.\notag\\
&\sum_{k\in [K]}W_k,\{Z_k\}_{k\in(\mathcal{T}_n\cup\{u\})\setminus \mathcal{S}_{[M]}}\Big) \notag\\
&-~I\big(\{X_k\}_{k \in [K]};\{W_k\}_{k\in\mathcal{S}_m} \big|\notag\\
&\sum_{k\in [K]}W_k, \{Z_k,W_k\}_{k\in(\mathcal{T}_n\cup\{u\})\setminus \mathcal{S}_{[M]}}\Big) \label{eq:tq}\\
\geq& H\big(\{X_k\}_{k \in \mathcal{U}_{m,n,u}\cap\mathcal{S}_{[M]}}\big|\{Z_k\}_{k\in(\mathcal{T}_n\cup\{u\})\setminus \mathcal{S}_{[M]}}\big) \notag\\
&-~I\big(\{X_k\}_{k \in [K]},\{W_k\}_{k\in(\mathcal{T}_n\cup\{u\})\setminus \mathcal{S}_{[M]}};\notag\\
& (W_{k})_{k\in(\mathcal{T}_n\cup\{u\}) \cap \mathcal{S}_{[M]}} \big|\sum_{k\in [K]}W_k,\{Z_k\}_{k\in(\mathcal{T}_n\cup\{u\})\setminus \mathcal{S}_{[M]}}\Big) \label{eq:lemm3t3}\\
\geq& \big|\mathcal{U}_{m,n,u}\cap\mathcal{S}_{[M]}\big| L \notag\\
&-~I\big(\{W_k\}_{k\in(\mathcal{T}_n\cup\{u\})\setminus \mathcal{S}_{[M]}}; (W_{k})_{k\in(\mathcal{T}_n\cup\{u\}) \cap \mathcal{S}_{[M]}} \big|\notag\\
&\sum_{k\in [K]}W_k,\{Z_k\}_{k\in(\mathcal{T}_n\cup\{u\})\setminus \mathcal{S}_{[M]}}\Big) \notag\\
&-~I\big(\{X_k\}_{k \in [K]}; (W_{k})_{k\in(\mathcal{T}_n\cup\{u\}) \cap \mathcal{S}_{[M]}} \big|\notag\\
&\sum_{k\in [K]}W_k,\{Z_k,W_k\}_{k\in(\mathcal{T}_n\cup\{u\})\setminus \mathcal{S}_{[M]}}\Big)
\label{eq:tq1}\\
\overset{(\ref{security})}{=}& \big|\mathcal{U}_{m,n,u}\cap \mathcal{S}_{[M]}\big| L,
\end{align}
where (\ref{eq:tq}) follows from expanding the mutual information using the chain rule and from the fact that $\mathcal{U}_{m,n,u}\cap \mathcal{S}_{[M]} = ((\mathcal{T}_n\cup\{u\})\cap \mathcal{S}_{[M]})\cup \mathcal{S}_m$ since $\mathcal{S}_m \cap (\mathcal{T}_n \cup \{u\}) = \emptyset$. The last term of (\ref{eq:tq}) is zero due to the security constraint (\ref{security}). In (\ref{eq:tq1}), the first term follows from the chain rule together with Lemma~\ref{lemma1} applied to each expanded term; the second term follows from the independence of $X_k$ and $Z_k$; and the third term is also zero as shown below.
\begin{align}
    &I\big(\{X_k\}_{k \in [K]}; (W_{i})_{k\in(\mathcal{T}_n\cup\{u\}) \cap \mathcal{S}_{[M]}} \big|\notag\\
    & \sum_{k\in [K]}W_k,\{Z_k,W_k\}_{k\in(\mathcal{T}_n\cup\{u\})\setminus \mathcal{S}_{[M]}}\Big)\\
    =&\sum_{j\in (\mathcal{T}_n\cup\{u\})\cap \mathcal{S}_{[M]}} I\Big(\{X_k\}_{k \in [K]}; W_{j}  \big|\sum_{k\in [K]}W_k, \notag\\
    &\{Z_k,W_k\}_{k\in(\mathcal{T}_n\cup\{u\})\setminus \mathcal{S}_{[M]}}, \{ W_{j'}\}_{j'\in (\mathcal{T}_n\cup\{u\}) \cap \mathcal{S}_{[M]}, j'<j}\big)\notag\\
    \leq &\sum_{j\in (\mathcal{T}_n\cup\{u\})\cap \mathcal{S}_{[M]}} I\left(\{X_k\}_{k \in [K]}, \{ Z_{j'}\}_{j'\in (\mathcal{T}_n\cup\{u\}) \cap \mathcal{S}_{[M]}, j'<j};\right.\notag\\
    &\left. W_{j}  \big|\sum_{k\in [K]}W_k,\{Z_k,W_k\}_{k\in(\mathcal{T}_n\cup\{u\})\setminus \mathcal{S}_{[M]}},\right. \notag\\
    &\{ W_{j'}\}_{j'\in (\mathcal{T}_n\cup\{u\}) \cap \mathcal{S}_{[M]}, j'<j}\big)\\  
    = &\sum_{j\in \mathcal{T}_n\cup\{u\}\cap \mathcal{S}_{[M]}} I(\{ Z_{j'}\}_{j'\in \mathcal{T}_n\cup\{u\} \cap \mathcal{S}_{[M]}, j'<j}; W_{j}  \big|\sum_{k\in [K]}W_k, \notag\\
    &\{Z_k,W_k\}_{k\in(\mathcal{T}_n\cup\{u\})\setminus \mathcal{S}_{[M]}}, \{ W_{j'}\}_{j'\in (\mathcal{T}_n\cup\{u\}) \cap \mathcal{S}_{[M]}, j'<j})\notag\\
    &-\sum_{j\in (\mathcal{T}_n\cup\{u\})\cap \mathcal{S}_{[M]}} I\Big(\{X_k\}_{k \in [K]}; W_{j}  \big|\sum_{k\in [K]}W_k, \notag\\
    &\{Z_k,W_k\}_{k\in(\mathcal{T}_n\cup\{u\})\setminus \mathcal{S}_{[M]}}, \{ W_{j'},Z_{j'}\}_{j'\in (\mathcal{T}_n\cup\{u\}) \cap \mathcal{S}_{[M]}, j'<j})\label{eq:lemm3tt1}\\ 
    =&0,
\end{align}
In (\ref{eq:lemm3tt1}), the first term is zero due to the independence between the inputs $W_k$ and the keys $Z_k$; the second term is zero because, under the security constraint~(\ref{security}), for user $u$, the security set $\mathcal{S}_m = \{W_j\}$ remains protected against any colluding users in $\mathcal{T}_{n'} \subset \mathcal{T}_n$.

Generalizing the above lemma to also include the implicit security input set $\mathcal{S}_I$, we have the following lemma.

\begin{lemma} \label{lemma4} 
For any $\mathcal{S}_m, \mathcal{T}_n \subseteq [K]$ and any $u\in[K]$, $m\in[M]$, $n\in[N]$
such that $\mathcal{S}_m \cap (\mathcal{T}_n \cup \{u\}) = \emptyset$ and 
$|\mathcal{S}_m \cup \mathcal{T}_n \cup \{u\}| \le K-1$, we have
\begin{align}
&H\big(\{Z_k\}_{k \in (\mathcal{S}_m\cup\mathcal{T}_n\cup\{u\})\cap\overline{\mathcal{S}}}\big|\{Z_k\}_{k\in(\mathcal{T}_n\cup\{u\})\setminus \overline{\mathcal{S}}}\big)\notag\\
\geq& \big|(\mathcal{S}_m\cup\mathcal{T}_n\cup\{u\})\cap\overline{\mathcal{S}}\big| L. \label{lemma4_eq}
\end{align}
\end{lemma}

\proof Recall that $\overline{\mathcal{S}} = \cup_{i\in [M]} \mathcal{S}_i \cup \mathcal{S}_I$.
\begin{align}
&H\big(\{Z_k\}_{k \in (\mathcal{S}_m\cup\mathcal{T}_n\cup\{u\})\cap(\cup_{i\in [M]} \mathcal{S}_i\cup\mathcal{S}_I)}\big|\{Z_k\}_{k\in(\mathcal{T}_n\cup\{u\})\setminus \overline{\mathcal{S}}}\big)\notag\\
=&H\big(\{Z_k\}_{k \in ((\mathcal{T}_n\cup\{u\})\cap\mathcal{S}_I)}\big|\{Z_k\}_{k\in(\mathcal{T}_n\cup\{u\})\setminus (\cup_{i\in [M]} \mathcal{S}_i\cup\mathcal{S}_I)}\big)\notag\\
&+H\big(\{Z_k\}_{k \in \left((\mathcal{S}_m\cup\mathcal{T}_n\cup\{u\})\cap(\cup_{i\in [M]} \mathcal{S}_i\cup\mathcal{S}_I)\right)\setminus((\mathcal{T}_n\cup\{u\})\cap\mathcal{S}_I)}\big|\notag\\
& \{Z_k\}_{k\in(\mathcal{T}_n\cup\{u\})\setminus \overline{\mathcal{S}}},\{Z_k\}_{k \in ((\mathcal{T}_n\cup\{u\})\cap\mathcal{S}_I)}\big) \\
\geq&|(\mathcal{T}_n\cup\{u\})\cap\mathcal{S}_I |L\notag\\
&+\big|\left((\mathcal{S}_m\cup\mathcal{T}_n\cup\{u\})\cap\overline{\mathcal{S}}\right)\setminus((\mathcal{T}_n\cup\{u\})\cap\mathcal{S}_I)\big|L \label{eq:tx} \\
=&|(\mathcal{T}_n\cup\{u\})\cap\mathcal{S}_I|L+\big|\left((\mathcal{S}_m\cup\mathcal{T}_n\cup\{u\})\cap\overline{\mathcal{S}}\right)\big|L\notag\\
&-|(\mathcal{S}_m\cup\mathcal{T}_n\cup\{u\})\cap\overline{\mathcal{S}}\cap((\mathcal{T}_n\cup\{u\})\cap\mathcal{S}_I)|L\\
=&|(\mathcal{T}_n\cup\{u\})\cap\mathcal{S}_I|L+\big|\left((\mathcal{S}_m\cup\mathcal{T}_n\cup\{u\})\cap\overline{\mathcal{S}}\right)\big|L\notag\\
&-|(\mathcal{T}_n\cup\{u\})\cap\mathcal{S}_I|L\\
=&\big|(\mathcal{S}_m\cup\mathcal{T}_n\cup\{u\})\cap\overline{\mathcal{S}}\big|L
\end{align}
where (\ref{eq:tx}) is derived as follows. First, consider the first term. For each element $u' \in (\mathcal{T}_n\cup \{u\}) \cap \mathcal{S}_I$, as $u' \in \mathcal{S}_I$, we have $\mathcal{S}_{m'}$ and $\mathcal{T}_{n'}$ such that $|\mathcal{S}_{m'} \cup \mathcal{T}_{n'}\cup \{u\}| = K-1$ and $\{u'\} \cup \mathcal{S}_{m'} \cup \mathcal{T}_{n'}\cup \{u\} = [K]$.
\begin{align}
&H\big(\{Z_k\}_{k \in (\mathcal{T}_n\cup\{u\})\cap\mathcal{S}_I}\big|\{Z_k\}_{k\in(\mathcal{T}_n\cup\{u\})\setminus (\cup_{i\in [M]} \mathcal{S}_i\cup\mathcal{S}_I)}\big) \notag\\
 \geq&\sum_{u': u'\in (\mathcal{T}_n\cup\{u\})\cap\mathcal{S}_I}H\big(Z_{u'}\big|\{Z_k\}_{k\in(\mathcal{T}_n\cup\{u\})\setminus (\cup_{i\in [M]} \mathcal{S}_i\cup\mathcal{S}_I)},\notag\\
 &\left.\{Z_k\}_{k\in ((\mathcal{T}_n\cup\{u\})\cap\mathcal{S}_I)\setminus\{u\}}\right) \\
 \geq&\sum_{u': u' \in (\mathcal{T}_n\cup\{u\})\cap\mathcal{S}_I}H\big(Z_{u'}\big|\{Z_k\}_{k\in[K]\setminus (\cup_{i\in [M]} \mathcal{S}_i\cup\{u\})}\big)\notag\\
 \geq&\sum_{u': u' \in (\mathcal{T}_n\cup\{u\})\cap\mathcal{S}_I}H\left(Z_{u'}\big|\{Z_k\}_{k\in[K]\setminus (\mathcal{S}_{m'}\cup\{u\})}\right)\\
 =&\sum_{u': u' \in (\mathcal{T}_n\cup\{u\})\cap\mathcal{S}_I}H\left(Z_{u'} \big|\{Z_k\}_{k\in(\mathcal{T}_{n'}\cup \{u\})}\right)\\
 \overset{(\ref{corollary1_eq})}{\geq}&|(\mathcal{T}_n\cup\{u\})\cap\mathcal{S}_I|L.
\end{align}
Second, consider the second term of (\ref{eq:tx}).
\begin{align}
    &H\big(\{Z_k\}_{k \in \left((\mathcal{S}_m\cup\mathcal{T}_n\cup\{u\})\cap(\cup_{i\in [M]} \mathcal{S}_i\cup\mathcal{S}_I)\right)\setminus((\mathcal{T}_n\cup\{u\})\cap\mathcal{S}_I)}\big|\notag\\
    &\{Z_k\}_{k\in(\mathcal{T}_n\cup\{u\})\setminus \overline{\mathcal{S}}},\{Z_k\}_{k \in ((\mathcal{T}_n\cup\{u\})\cap\mathcal{S}_I)}\big) \notag\\
    =&H\big(\{Z_k\}_{k \in \left((\mathcal{S}_m\cup\mathcal{T}_n\cup\{u\})\cap(\cup_{i\in [M]} \mathcal{S}_i\cup\mathcal{S}_I)\right)\setminus((\mathcal{T}_n\cup\{u\})\cap\mathcal{S}_I)}\big|\notag\\
    &\{Z_k\}_{k\in(\mathcal{T}_n\cup\{u\})\setminus (\cup_{i\in [M]} \mathcal{S}_i\cup\mathcal{S}_I)},\{Z_k\}_{k \in ((\mathcal{T}_n\cup\{u\})\cap\mathcal{S}_I)}\big) \notag\\
    =&H\big(\{Z_k\}_{k \in \left((\mathcal{S}_m\cup\mathcal{T}_n\cup\{u\})\cap(\cup_{i\in [M]} \mathcal{S}_i\cup\mathcal{S}_I)\right)\setminus((\mathcal{T}_n\cup\{u\})\cap\mathcal{S}_I)}\big|\notag\\
    &\{Z_k\}_{k\in(\mathcal{T}_n\cup\{u\})\setminus (\cup_{i\in [M]} \mathcal{S}_i)}\big) \\
    \overset{(\ref{lemma3_eq})}{\geq}&\big|\left((\mathcal{S}_m\cup\mathcal{T}_n\cup\{u\})\cap\overline{\mathcal{S}}\right)\setminus((\mathcal{T}_n\cup\{u\})\cap\mathcal{S}_I)\big|L
    \label{lemma4_eq3}
\end{align}
where (\ref{lemma4_eq3}) follows from the fact that $\big((\mathcal{S}_m\cup\mathcal{T}_n\cup\{u\})\cap$ $(\cup_{i\in [M]} \mathcal{S}_i\cup\mathcal{S}_I)\big)\setminus$ $((\mathcal{T}_n\cup \{u\})\cap\mathcal{S}_I)= (\mathcal{S}_m\cup\mathcal{T}_n\cup\{u\})\cap(\cup_{i\in [M]} \mathcal{S}_i)$.


We first establish that $R_X \geq 1$, which holds in all cases.
By Lemma~\ref{lemma1} we have
\begin{align}
    &H\left(X_{u}\right)\geq H\left(X_u\big|\{W_k,Z_k\}_{k\in[K]\backslash \{u\}}\right) \overset{(\ref{lemma1_eq})}{\geq} L, \label{lemma1_eqXX}\\
    \Rightarrow&R_X= \frac{L_X}{L}\geq \frac{H(X_{u})}{L}\geq 1.
\end{align}


Consider the `If' case $a^* = K$, where there exist $\mathcal{S}_m$, $\mathcal{T}_n$, and a user $u$ such that
$
|\mathcal{S}_m \cup \mathcal{T}_n \cup \{u\}| = K.
$
Since the set $\{\mathcal{S}_m\}_{m\in[M]}$ and $\{\mathcal{T}_n\}_{n\in[N]}$ are monotone, we may select $\mathcal{S}_{m'}$, $\mathcal{T}_n$, and the same user $u$ such that
$
|\mathcal{S}_{m'} \cup \mathcal{T}_n \cup \{u\}| = K-1,
|(\mathcal{S}_{m'} \cup \mathcal{T}_n \cup \{u\}) \cap \overline{\mathcal{S}}| = K-1.
$
Applying Lemma~\ref{lemma4}, we have
\begin{align}
H(Z_{\Sigma})
&\overset{(\ref{total rand})}{\ge} H\left(\{Z_k\}_{k\in[K]}\right) \\
&\ge H\big(\{Z_k\}_{k \in (\mathcal{S}_{m'}\cup\mathcal{T}_n\cup\{u\})\cap\overline{\mathcal{S}}}
\,\big|
\{Z_k\}_{k \in (\mathcal{T}_n\cup\{u\})\setminus\overline{\mathcal{S}}}
\big) \notag\\
&\overset{(\ref{lemma4_eq})}{\ge} \big|(\mathcal{S}_{m'}\cup\mathcal{T}_n\cup\{u\})\cap\overline{\mathcal{S}}\big| L
= (K-1)L.
\end{align}
Hence
$
R_{Z_{\Sigma}} = \frac{L_{Z_{\Sigma}}}{L}
\ge \frac{H(Z_{\Sigma})}{L}
\ge K-1.
$


Consider the `else if' case $a^* \leq K-1$,
We are now ready to show that $R_{Z_\Sigma} \ge a^*$. 
To proceed, consider any $\mathcal{A}_{m,n,u}$ whose $\mathcal{S}_m$, $\mathcal{T}_n$, and $u$ satisfy the conditions in Lemma~\ref{lemma4}. 
Since the set systems are monotone, we may, without loss of generality, assume that $\mathcal{S}_m \cap (\mathcal{T}_n \cup\{u\}) = \emptyset$.
\begin{align}
    H(Z_{\Sigma})\overset{(\ref{total rand})}{\geq}&H\left(\{Z_k\}_{k\in [K]}\right)\\
    \geq&H\big(\{Z_k\}_{k \in (\mathcal{S}_{m}\cup\mathcal{T}_{n}\cup \{u\})\cap\overline{\mathcal{S}}}\big|\{Z_k\}_{k\in(\mathcal{T}_{n}\cup \{u\})\setminus \overline{\mathcal{S}}}\big) \notag\\
    \overset{(\ref{lemma4_eq})}{\geq}& \big|(\mathcal{S}_{m}\cup\mathcal{T}_{n}\cup \{u\})\cap\overline{\mathcal{S}}\big| L = | \mathcal{A}_{m,n,u}| L\\
\Rightarrow ~~&H(Z_{\Sigma}) \geq \mbox{max}_{m,n,u} |\mathcal{A}_{m,n,u}| L =a^*L\label{totalrand_eq1}\\
\Rightarrow ~~&R_{Z_{\Sigma}} = \frac{L_{Z_{\Sigma}}}{L} \geq \frac{H(Z_{\Sigma})}{L} \geq a^*.
\end{align}

The converse proof for the `If' and `else if' cases is now complete.

\subsection{Converse Proof of Example \ref{ex2}}
Next, we consider the `otherwise' case of Theorem \ref{thm:d1}, which requires an additional step to further tighten the bound on $R_{Z_\Sigma}$ from $a^*$ to $a^* + b^*$. In this case, we have $a^* \le K-1$. To illustrate the idea more clearly, we begin with a simple example.

\begin{example}\label{ex2}
Consider $K=6$, the security input sets are $(\mathcal{S}_1,\mathcal{S}_2,\mathcal{S}_3) =(\emptyset, \{1\}, \{2\})$, and the colluding user sets are $(\mathcal{T}_1,\cdots,\mathcal{T}_{11})=(\emptyset, \{1\}, \{2\}, \{3\}, \{4\},$ $\{5\},$ $\{6\},$ $ \{1,3\},$ $ \{2,4\},$ $ \{2,5\}, \{1,6\})$. 
\end{example}

For Example \ref{ex2}, the implicit security set $\mathcal{S}_I$ is empty, and $\overline{\mathcal{S}}= \cup_{m} \mathcal{S}_m =\{1,2\}$. Finding all $\mathcal{A}_{m,n,u}$ with maximum cardinality, we have $\mathcal{A}_{2,9,3}  =\mathcal{A}_{2,9,5}  =\mathcal{A}_{2,9,6}  =\mathcal{A}_{2,10,3}  =\mathcal{A}_{2,10,4}  =\mathcal{A}_{2,10,6}  =\mathcal{A}_{3,8,4}  =\mathcal{A}_{3,8,5}  =\mathcal{A}_{3,8,6}  =\mathcal{A}_{3,11,3}  =\mathcal{A}_{3,11,4}  =\mathcal{A}_{3,11,5}  = \overline{\mathcal{S}} =  \{1,2\}$, and $a^* = \big|\overline{\mathcal{S}}\big| = 2 \leq K-1 = 4$. Then $\mathcal{Q} = \cup_{m,n,u: |\mathcal{A}_{m,n,u}| = a^*} \mathcal{S}_m \cup \mathcal{T}_n\cup\{u\} = \{1,2,3,4,5,6\}$ and $|\mathcal{Q}| = 6 = K$. So we are in the `otherwise' case.

Consider all $\mathcal{A}_{m,n,u}$ sets so that $|\mathcal{A}_{m,n,u}| = a^*$. For Example \ref{ex2}, we have $9$ such sets $\mathcal{A}_{2,9,3},$ $\mathcal{A}_{2,9,5},$ $ \mathcal{A}_{2,9,6},$ $ \mathcal{A}_{2,10,3}, \mathcal{A}_{2,10,4},$ $ \mathcal{A}_{2,10,6},$ $ \mathcal{A}_{3,8,4}, \mathcal{A}_{3,8,5},$ $ \mathcal{A}_{3,8,6},$ $ \mathcal{A}_{3,11,3},$ $ \mathcal{A}_{3,11,4},$ $ \mathcal{A}_{3,11,5}$. Note that $\mathcal{S}_m \cup \mathcal{T}_n\cup \{u\} = \mathcal{A}_{m,n,u} \cup ((\mathcal{T}_n\cup\{u\}) \setminus \overline{\mathcal{S}})$, and we consider the key variables $Z_k$ in the set $\mathcal{S}_m \cup \mathcal{T}_n \cup\{u\}$ and split them to $\mathcal{A}_{m,n,u}$ (treated by Lemma \ref{lemma4} and corresponds to $a^*$) and $\mathcal{T}_n\cup \{u\} \setminus \overline{\mathcal{S}}$ (the new part treated by a linear program and corresponds to $b^*$).
\begin{align}
&H(Z_\Sigma)\notag\\
\geq& \max\big( H( \left\{Z_k\right\}_{k \in \mathcal{S}_2 \cup \mathcal{T}_9\cup \{3\}} ), H( \left\{Z_k\right\}_{k \in \mathcal{S}_2 \cup \mathcal{T}_9\cup \{5\}} ),\notag\\
&~~~~~~~H( \left\{Z_k\right\}_{k \in \mathcal{S}_2 \cup \mathcal{T}_9\cup \{6\}} ), H( \left\{Z_k\right\}_{k \in \mathcal{S}_2 \cup \mathcal{T}_{10}\cup \{3\}} ), \notag\\
&~~~~~~~H( \left\{Z_k\right\}_{k \in \mathcal{S}_2 \cup \mathcal{T}_{10}\cup \{4\}} ), H( \left\{Z_k\right\}_{k \in \mathcal{S}_2 \cup \mathcal{T}_{10}\cup \{6\}} ), \notag\\
&~~~~~~~H( \left\{Z_k\right\}_{k \in \mathcal{S}_2 \cup \mathcal{T}_{8}\cup \{4\}} ), H( \left\{Z_k\right\}_{k \in \mathcal{S}_2 \cup \mathcal{T}_{8}\cup \{5\}} ), \notag\\
&~~~~~~~H( \left\{Z_k\right\}_{k \in \mathcal{S}_2 \cup \mathcal{T}_8\cup \{6\}} ), H( \left\{Z_k\right\}_{k \in \mathcal{S}_2 \cup \mathcal{T}_{10}\cup \{3\}} ), \notag\\
&~~~~~~~H( \left\{Z_k\right\}_{k \in \mathcal{S}_3 \cup \mathcal{T}_{11}\cup \{4\}} ), H( \left\{Z_k\right\}_{k \in \mathcal{S}_3 \cup \mathcal{T}_{11}\cup \{5\}} )\big) \notag\\
=& \max \big( H(Z_1, Z_2,Z_3,Z_4), H(Z_1, Z_2,Z_4,Z_5),\notag\\
&~~~~~~~H(Z_1, Z_2,Z_4,Z_6), H(Z_1, Z_2,Z_3,Z_5),   \notag\\
&~~~~~~~ H(Z_1,Z_2,Z_3,Z_6), H(Z_1, Z_2,Z_5,Z_6)\big)\label{eq:tt}\\
\overset{(\ref{lemma4_eq})}{\geq}& \max \big( H(Z_3,Z_4), H(Z_4,Z_5), H(Z_4,Z_6),\notag\\
&~~~~~~~ H(Z_3,Z_5), H(Z_3,Z_6), H(Z_5,Z_6)\big)+ a^*L \label{eq:tt1}
\end{align}
In (\ref{eq:tt1}), we split the non-redundant $Z_k$ terms in the set $\mathcal{S}_m \cup \mathcal{T}_n \cup \{u\}$ into two parts: those in $(\mathcal{T}_n \cup \{u\}) \setminus \overline{\mathcal{S}}$ and those in $\mathcal{A}_{m,n,u}$. We then apply Lemma \ref{lemma4} to bound the $\mathcal{A}_{m,n,u}$ term conditioned on $(\mathcal{T}_n \cup \{u\}) \setminus \overline{\mathcal{S}}$. Note that we consider only the case where $|\mathcal{A}_{m,n,u}| = a^*$, i.e., $\mathcal{A}_{m,n,u} = \overline{\mathcal{S}}$.
Next, we proceed to bound the term $\max(H(Z_3,Z_4), H(Z_4,Z_5), H(Z_4,Z_6), H(Z_3,Z_5),$ $ H(Z_3,Z_6),$ $ H(Z_5,Z_6) )$, where it turns out that the only constraints required are from Lemma \ref{lemma2}. For example, consider $\mathcal{S}_2$, $\mathcal{T}_9$, and user~3. From~(\ref{lemma2_eq}) we have $H(Z_5,Z_6 \mid Z_2,Z_3,Z_4) \ge L$. Similarly, for any $\mathcal{S}_m$, $\mathcal{T}_n$, and user~$u$ such that $|\mathcal{A}_{m,n,u}| = a^*$, we have 
\begin{align}
   & H(Z_5,Z_6|Z_2,Z_3,Z_4)\geq L,  H(Z_3,Z_6|Z_2,Z_4,Z_5)\geq L, \notag \\
    & H(Z_3,Z_5|Z_2,Z_4,Z_6)\geq L,  H(Z_4,Z_6|Z_2,Z_3,Z_5)\geq L, \notag \\
    & H(Z_3,Z_6|Z_2,Z_4,Z_5)\geq L,  H(Z_3,Z_4|Z_2,Z_5,Z_6)\geq L, \notag \\
    & H(Z_5,Z_6|Z_1,Z_3,Z_4)\geq L,  H(Z_4,Z_6|Z_1,Z_3,Z_5)\geq L, \notag \\
    & H(Z_4,Z_5|Z_1,Z_3,Z_6)\geq L,  H(Z_4,Z_5|Z_1,Z_3,Z_6)\geq L, \notag \\
    & H(Z_3,Z_5|Z_1,Z_4,Z_6)\geq L,  H(Z_3,Z_4|Z_1,Z_5,Z_6)\geq L. \label{eq:tt3}
\end{align}
To bound (\ref{eq:tt1}) under the constraints in (\ref{eq:tt3}), a linear program is formulated in which the conditional entropy terms serve as variables consistent with the chain-rule expansion. In particular, set
\begin{align}
&H(Z_3) = b_3L,\quad H(Z_4|Z_3) = b_4L, \notag\\
&H(Z_5|Z_3,Z_4) = b_5L,\quad H(Z_6|Z_3,Z_4,Z_5) = b_6L. \label{setlinear}
\end{align}
Substituting (\ref{eq:tt1}) and (\ref{setlinear}) then gives
\begin{align}
R_{Z_\Sigma} \geq& a^* + \min \max(b_3+b_4, b_4+b_5, b_4+b_6, \notag\\
&~~~~~~~~~~~~~~~~~~ b_3+b_5, b_3+b_6, b_5+b_6), \label{minlp}
\end{align}
where the $\min$ operator in (\ref{minlp}) is used to obtain the tightest possible lower bound.
The minimization is taken over the following linear constraints, which follow from (\ref{eq:tt3}) and the definitions in (\ref{setlinear}):
{\setlength{\arraycolsep}{1pt}
\begin{align}
   & b_5+b_6 \geq 1,~~ b_3+b_6 \geq 1,~~ b_3+b_5 \geq 1,~~ b_4+b_6 \geq 1, \notag \\
   &b_3+b_6 \geq 1,~~ b_3+b_4 \geq 1,~~b_5+b_6 \geq 1,~~ b_4+b_6 \geq 1, \notag \\
   &b_4+b_5\geq 1,~~ b_4+b_5 \geq 1,~~ b_3+b_5 \geq 1,~~b_3+b_4 \geq 1, \notag \\
   & b_3, b_4, b_5,b_6 \geq 0. \label{eq:tt4}
\end{align}}
For the first inequality, $b_5 + b_6 \ge 1$, we have $b_5 + b_6 \ge (H(Z_5 \mid Z_2, Z_3, Z_4) + H(Z_6 \mid Z_2, Z_3, Z_4, Z_5))/L \ge 1$, and the remaining inequalities follow similarly. The correlations among $H(Z_3), H(Z_4), H(Z_5), H(Z_6)$ are captured by (\ref{eq:tt3}), and the tightest bound under these constraints can be obtained via a linear program over non-negative variables $b_3, b_4, b_5, b_6$, where each maximal $\mathcal{A}_{m,n,u}$ set contributes one constraint (redundancies removed in (\ref{eq:tt4})). Defining $b^* = \min \max(b_3+b_4, b_4+b_5, b_4+b_6, b_3+b_5, b_3+b_6, b_5+b_6)$, the converse is then $R_{Z_\Sigma} \ge a^* + b^*$. For Example \ref{ex2}, the optimum is $b^* = 1$, achieved when $b_3 = b_4 = b_5 = b_6 = 1/2$, which also informs the achievability proof in Section \ref{sec:ach}.

\subsection{Converse Proof of $R_{Z_\Sigma} \geq a^* + b^*$}
Extending the insights from Example \ref{ex2}, we now provide the general proof that $R_{Z_\Sigma} \geq a^* + b^*$ under the `otherwise' case, where $a^* \le K-1$, $a^* = |\overline{\mathcal{S}}|$, and $|\mathcal{Q}| = K$.

Consider all $\mathcal{S}_m, \mathcal{T}*n$ and user $u$ such that $|\mathcal{A}_{m,n,u}| = a^*$. In the `otherwise' case, we have $a^* = |\overline{\mathcal{S}}|$ and hence $\mathcal{A}_{m,n,u} = \overline{\mathcal{S}}$. Note that
$\mathcal{S}_m \cup \mathcal{T}_n \cup \{u\} = \mathcal{A}_{m,n,u} \cup \big((\mathcal{T}_n \cup \{u\}) \setminus \overline{\mathcal{S}}\big),$
so the key variables $Z_k$ in this set can be split into two parts: $\mathcal{A}_{m,n,u}$, handled by Lemma~\ref{lemma4} (contributing $a^*$), and $(\mathcal{T}_n \cup \{u\}) \setminus \overline{\mathcal{S}}$, handled via a linear program (contributing $b^*$).
{\setlength{\arraycolsep}{1pt}
\begin{align}
  &H(Z_{\Sigma})\notag\\
  \overset{(\ref{total rand})}
  {\geq}& \max_{m,n,u: |\mathcal{A}_{m,n,u}| = a^*}  H\big((Z_k)_{k\in \mathcal{S}_m \cup \mathcal{T}_n\cup\{u\}}\big) \\
    \geq&\max_{m,n,u: |\mathcal{A}_{m,n,u}| = a^*}  \big( H\big((Z_k)_{k \in (\mathcal{T}_{n}\cup\{u\})\setminus\overline{\mathcal{S}}}\big) 
    \notag\\
    &+H\big((Z_k)_{k \in (\mathcal{S}_{m}\cup\mathcal{T}_{n}\cup\{u\})\cap\overline{\mathcal{S}}}\big|(Z_k)_{k\in(\mathcal{T}_{n}\cup\{u\})\setminus \overline{\mathcal{S}}}\big)  \big) \\
    \overset{(\ref{lemma4_eq})}{\geq}& \max_{m,n,u: |\mathcal{A}_{m,n,u}| = a^*}  H\big((Z_k)_{k\in (\mathcal{T}_{n}\cup\{u\})\setminus\overline{\mathcal{S}}}\big)  + a^*L \label{eq:con1} 
\end{align}
}
subject to the following constraints by Lemma \ref{lemma2},
{\setlength{\arraycolsep}{1pt}
\begin{align}
&\forall m,n,u ~\mbox{where $|\mathcal{A}_{m,n,u}| = a^*$}:\notag\\
&H\left((Z_k)_{k \in [K] \setminus(\mathcal{S}_{m}\cup\mathcal{T}_{n}\cup\{u\})}\big|(Z_k)_{k\in(\mathcal{T}_{n}\cup\{u\})}\right) &\overset{(\ref{lemma2_eq})}{\geq}& L. \label{eq:con2}
\end{align}
}
Next, following the proof steps in Example~\ref{ex2}, we convert inequality~(\ref{eq:con1}) together with the constraints in~(\ref{eq:con2}) into a linear program in terms of the variables $b_k$, defined as follows.
\begin{eqnarray}
b_k \triangleq H\big(Z_{k}\big|(Z_{l})_{l\in[K]\setminus\overline{\mathcal{S}}, l < k}\big)/L, ~~\forall k\in [K]\setminus\overline{\mathcal{S}}. \label{assume_eq1}
\end{eqnarray}
Then, by normalizing (\ref{eq:con1}) with respect to $L$ on both sides and expanding the entropy terms in (\ref{eq:con1}) and (\ref{eq:con2}) using the chain rule under the lexicographic order, we obtain the following linear program.
\begin{align}
&R_{Z_\Sigma} \geq a^* + \min \max_{m,n,u: |\mathcal{A}_{m,n,u}| = a^*} \Bigg( \sum_{k\in (\mathcal{T}_n\cup\{u\}) \setminus \overline{\mathcal{S}} } b_k\Bigg)\label{lpcoverser1}\\
&\mbox{subject to}\notag\\
&\sum_{k \in [K] \setminus (\mathcal{S}_m \cup \mathcal{T}_n\cup \{u\})} b_k \geq 1, \forall m,n,u ~\mbox{such that}~|\mathcal{A}_{m,n,u}| = a^*, \label{lpcoverser2}\\
&~~~~~~~ b_k \geq 0, ~~~~~~k \in [K]\setminus \overline{\mathcal{S}}. \label{lpcoverser}
\end{align}
Note that $(\mathcal{S}_m \cup \mathcal{T}_n \cup \{u\}) \cap \overline{\mathcal{S}} = \mathcal{A}_{m,n,u} = \overline{\mathcal{S}} \subset (\mathcal{S}_m \cup \mathcal{T}_n \cup \{u\})$, so the chain-rule expansion above can be lower-bounded by the corresponding $b_k$ terms. Based on the linear program (\ref{lp}), whose optimal value is denoted by $b^*$, we thereby obtain the desired converse bound $R_{Z_\Sigma} \geq a^* + b^*$.

\section{Achievability Proof of Theorem \ref{thm:d1}} \label{sec:ach}
\subsection{Achievable Scheme  for `If' Case where $a^* = K$}\label{sec:ach111}
In this case $a^{*}=K$. We assign each user one key variable. By applying the scheme of Theorem~1 in \cite{Zhang_Li_Wan_DSA}, we can achieve the communication rate $R_X=1$ and the source key rate $R_{Z_\Sigma}=K-1$.


\subsection{Achievable Scheme for Subcase 1 of the `Else if' Case where $a^* < \big|\overline{\mathcal{S}}\big|$}\label{sec:ach1}
In this case, each user in the (implicit or explicit) security input set $\overline{\mathcal{S}}$ is assigned a key variable, whereas users in $[K]\setminus\overline{\mathcal{S}}$ are assigned none. We work over the field $\mathbb{F}_{q}$ and choose $q > a^* \binom{|\overline{\mathcal{S}}|}{a^*}$ to satisfy the required security constraint. Let $Z_\Sigma = {\bf N} = (N_1;\ldots;N_{a^*}) \in \mathbb{F}_{q}^{a^* \times 1}$ denote a vector of $a^*$ i.i.d.\ uniform variables. The key variables are generated as 
\begin{eqnarray}
 Z_k &=& {\bf H}_k\times {\bf N}, \quad \forall k \in \overline{\mathcal{S}},\notag\\
 Z_k &=& 0, \forall k \in [K]\setminus\overline{\mathcal{S}},\label{eq:c11}
\end{eqnarray}
where ${\bf H}_k \in \mathbb{F}_{q}^{1\times a^*}$. Without loss of generality, let $\overline{\mathcal{S}} = \{k_1, \ldots, k_{|\overline{\mathcal{S}}|}\}$.
To construct the coefficient vectors, we draw each element of ${\bf H}_{k_1}, \ldots, {\bf H}_{k_{|\overline{\mathcal{S}}|-1}}$ independently and uniformly from $\mathbb{F}_{q}$. For correctness, the final vector is defined as the additive inverse of their sum:
\begin{align}
    {\bf H}_{k_{|\overline{\mathcal{S}}|}}
    \triangleq
    -\left( {\bf H}_{k_1} + \cdots + {\bf H}_{k_{|\overline{\mathcal{S}}|-1}} \right).
    \label{eq:c12}
\end{align}
There must exist a realization\footnote{Since $a^* < |\overline{\mathcal{S}}|$, the determinant associated with any selection of $a^*$ distinct vectors ${\bf h}_k$ defines a non-zero polynomial. Applying the Schwartz--Zippel lemma, the product of all such determinant polynomials has degree $a^* \binom{|\overline{\mathcal{S}}|}{a^*}$. By choosing a field $\mathbb{F}_q$ with $q$ larger than this degree, the probability that the product polynomial evaluates to a non-zero value is strictly positive. This ensures the validity of (\ref{eq:sz1}).
} of ${\bf H}_k, k \in \overline{\mathcal{S}}$ such that
\begin{align}
&\mbox{Any subset of at most $a^*$ distinct vectors ${\bf H}_k$, $k \in \overline{\mathcal{S}}$, } \notag\\
&\mbox{is linearly independent.} \label{eq:sz1}
\end{align}
Moreover,
\begin{eqnarray}
 \sum_{k\in\overline{\mathcal{S}}} {\bf H}_k = {\bf 0}
 ~~\Rightarrow~~
 \sum_{k\in[K]} Z_k 
 \overset{(\ref{eq:c11})}{=}
 \sum_{k\in\overline{\mathcal{S}}} Z_k
 = 0. \label{eq:c13}
\end{eqnarray}
We set $L_X = L = 1$ and define the transmitted messages as
\begin{eqnarray}
     X_k = W_k + Z_k,\quad \forall k \in [K].\label{eq:m1}
\end{eqnarray}
The achieved communication rate is $R_X = L_X / L = 1$, and the source key rate is $R_{Z_\Sigma} = L_{Z_\Sigma} / L = a^*$, as desired. For correctness, consider any user $u \in [K]$. We have
$\sum_{k \in [K]\setminus\{u\}} X_k + W_u + Z_u = \sum_{k \in [K]\setminus\{u\}} X_k \overset{(\ref{eq:c13})}{=} \sum_{k \in [K]} W_k,$
which ensures that each user can recover the desired sum. The security proof is provided in Section~\ref{sec:security}, where we show that a realization of the randomly generated vectors ${\bf H}_k$ exists that satisfies the required full-rank property.


\subsection{Achievable Scheme for Subcase 2 of the `Else if' Case where  $a^* = \big|\overline{\mathcal{S}}\big|$, $|\mathcal{Q}| < K$}\label{sec:ach2}
In this case, every user in the total security input set $\overline{\mathcal{S}}$ and one user outside $\mathcal{Q}$ will be assigned a key variable.
 Pick $q$ so that $q > a^* \binom{|\overline{\mathcal{S}}|+1}{a^*} = a^*(a^*+1)$ and operate over $\mathbb{F}_{q}$.
 Consider $a^*$ i.i.d. uniform variables, $Z_\Sigma ={\bf N} = (N_1;\cdots;N_{a^*}) \in \mathbb{F}_{q}^{a^* \times 1}$. Find any $u' \in [K]\setminus\mathcal{Q}$ (which must exist as $|\mathcal{Q}| < K$ and $u' \notin \overline{\mathcal{S}}$ as $a^* = \big|\overline{\mathcal{S}}\big|$ and $\overline{\mathcal{S}} \subset \mathcal{Q}$)
 and set the key variables as 
 \begin{eqnarray}
 Z_k &=& {\bf H}_k \times {\bf N}, \forall k \in (\overline{\mathcal{S}} \cup \{u'\}), \notag\\
 Z_k &=& 0, \forall k \in [K]\setminus(\overline{\mathcal{S}}\cup\{u'\}),\label{eq:c21}
 \end{eqnarray}
 where ${\bf H}_k \in \mathbb{F}_{q}^{1\times a^*}$. Without loss of generality, let $\overline{\mathcal{S}} = \{k_1, \ldots, k_{|\overline{\mathcal{S}}|}\}$.
To construct the coefficient vectors, we draw each element of ${\bf H}_{k_1}, \ldots, {\bf H}_{k_{|\overline{\mathcal{S}}|}}$ independently and uniformly from $\mathbb{F}_{q}$. For correctness, the final vector is defined as the additive inverse of their sum:
\begin{align}
    {\bf H}_{u'}
    \triangleq
    -\left( {\bf H}_{k_1} + \cdots + {\bf H}_{k_{|\overline{\mathcal{S}}|}} \right).
    \label{eq:c22}
\end{align}
Using the same argument based on the Schwartz--Zippel lemma, one can choose a realization 
of ${\bf H}_k$, $k \in \overline{\mathcal{S}} \cup \{u'\}$, such that
\begin{align}
&\mbox{Any subset of at most $a^*$ distinct vectors ${\bf H}_k$, $k \in \overline{\mathcal{S}} \cup \{u'\}$,} \notag\\
&\mbox{is linearly independent.} \label{eq:sz2}
\end{align}
Moreover,
\begin{eqnarray}
 &&\sum_{k\in(\overline{\mathcal{S}}\cup\{u'\})} {\bf H}_k = {\bf 0}\notag\\
 \Rightarrow&&
 \sum_{k\in[K]} Z_k 
 \overset{(\ref{eq:c11})}{=}
 \sum_{k\in(\overline{\mathcal{S}}\cup\{u'\})} Z_k = {\bf 0}.~ \label{eq:c23}
\end{eqnarray}
Finally, set $L_X=L=1$ and the sent messages as
 \begin{eqnarray}
     X_k = W_k + Z_k, \forall k \in [K].\label{eq:m2}
 \end{eqnarray}
 The achieved communication rate is $R_X = L_X / L = 1$, and the source key rate is $R_{Z_\Sigma} = L_{Z_\Sigma} / L = a^*$, as desired. For correctness, consider any user $u \in [K]$. We have
$\sum_{k \in [K]\setminus\{u\}} X_k + W_u + Z_u = \sum_{k \in [K]\setminus\{u\}} X_k \overset{(\ref{eq:c13})}{=} \sum_{k \in [K]} W_k,$
which ensures that each user can recover the desired sum. The security proof is presented in a unified manner in Section \ref{sec:security}. 

\subsection{Achievable Scheme of Example \ref{ex2}}
Before presenting the general achievable scheme that attains $R_{Z_\Sigma} = a^* + b^*$ in the ``Otherwise'' case, this section revisits Example~\ref{ex2} and provides an explicit coding construction over small finite fields.

In this example, we have $a^*=2$ and $b^*=2/2$, so $R_{Z_\Sigma}=L_{Z_\Sigma}/L=6/2$ is achievable. Let $q=5$, meaning that all symbols are defined over $\mathbb{F}_{5}$. Consider 6 i.i.d. uniform variables $(N_1,N_2,N_3,N_4,N_5,N_6)=Z_\Sigma$ in $\mathbb{F}_{5}$, so $L_{Z_\Sigma}=6$. Set $L=2$, and represent each user input as $W_k=(W_{k,1},W_{k,2})$ for all $k\in[6]$. For users $1$ and $2$ in $\overline{\mathcal{S}}$, assign a key of size $L=2$ symbols. For users $3,4,5,6$ in $[K]\setminus\overline{\mathcal{S}}$, the key sizes follow from the linear program in (\ref{setlinear})(\ref{eq:tt4}), giving $L_{Z_3}=L_{Z_4}=L_{Z_5}=L_{Z_6}=1$. Define the key variables as follows:
\begin{align}
& Z_1 = \left[
\begin{array}{c}
 - N_1 - N_3 - N_4 - N_5 - N_6\\
- N_2 - N_3 - 2N_4 - 3N_5 - 4N_6
\end{array}
\right],
Z_2 = \left[
\begin{array}{c}
N_1 \\
N_2 
\end{array}
\right],\notag\\
&Z_3 = N_3,~Z_4 = N_4,~Z_5 = N_5,~Z_6 = N_6. 
\end{align}
Finally, set the message $X_k$ as 
\begin{align}
& X_1 = \left[
\begin{array}{c}
W_{1,1} - N_1 - N_3 - N_4 - N_5 - N_6\\
W_{1,2} - N_2 - N_3 - 2N_4 - 3N_5 - 4N_6
\end{array}
\right],\notag
\\
&
X_2 = \left[
\begin{array}{c}
W_{2,1} + N_1 \\
W_{2,2} + N_2 
\end{array}
\right],~
X_3 = \left[
\begin{array}{c}
W_{3,1} + N_3 \\
W_{3,2} + N_3 
\end{array}
\right],\notag\\
& 
X_4 = \left[
\begin{array}{c}
W_{4,1} + N_4 \\
W_{4,2} + 2N_4 
\end{array}
\right], ~
X_5 = \left[
\begin{array}{c}
W_{5,1} + N_5 \\
W_{5,2} + 3N_5 
\end{array}
\right],\notag\\
&X_6 = \left[
\begin{array}{c}
W_{6,1} + N_6 \\
W_{6,2} + 4N_6 
\end{array}
\right].
\end{align} 

For correctness, consider any user $u \in [6]$. We have
$\sum_{k \in [6]\setminus\{u\}} X_k + W_u + Z_u = \sum_{k \in [6]\setminus\{u\}} X_k = \sum_{k \in [K]} W_k,$
which ensures that each user can recover the desired sum. For security, we verify one case of $\mathcal{S}_2 = \{1\}, \mathcal{T}_9 = \{2,4\}, \{u\}=\{3\}$; other cases are similar and are deferred to the general proof in Section \ref{sec:security}. 
\begin{align}
& I\Big( W_1 ; \{X_k\}_{k\in [6]} \Big| \sum_{k\in [6]}W_k, W_2, N_1, N_2,W_3, N_3, W_4, N_4 \Big) \notag\\
\leq& H(X_1, X_5, X_6 | W_1 + W_5 + W_6, N_1, N_2, N_3, N_4)- \notag\\
& H(X_1, X_5, X_6 | W_1, W_5+W_6, W_2, N_1, N_2,W_3, N_3, W_4, N_4) \notag\\
\leq& 4 - H(S_5, S_6, W_5, W_6 | W_5 + W_6) \label{tmp} \\
=& 4 - 4 = 0
\end{align}
where the first term of (\ref{tmp}) is due to the fact that $X_1, X_5, X_6$ contain $4$ variables given the conditioning terms and we have repeatedly used the independence of the randomness variables and input variables. Thus the security constraint (\ref{security}) is satisfied.


The communication rate achieved is $R_X=L_X/L=2/2$, and the key rate achieved is $R_{Z_\Sigma} = L_Z/L =6/2$, as desired.

\subsection{Achievable Scheme of for `Otherwise' Case where  $a^* = \big|\overline{\mathcal{S}}\big|$, $|\mathcal{Q}| = K$}\label{sec:ach3}
In the `Otherwise' case, every user is assigned a key variable. For users in $\overline{\mathcal{S}}$ assign a key of size $L$ symbols. For users in $[K]\setminus\overline{\mathcal{S}}$ assign key sizes according to the optimal solution of the linear program in (\ref{lpcoverser1})(\ref{lpcoverser2})(\ref{lpcoverser}), namely $L_{Z_k}=b_k^* L$ for $k\in[K]\setminus\overline{\mathcal{S}}$. Denote the values $b_k$ that attain the optimal objective $b^*$ of (\ref{lp}) by
\begin{equation}
    b_k = b_k^* = \frac{p_k}{\overline{q}}, \qquad \forall k\in [K]\setminus\overline{\mathcal{S}}. \label{eq:bk1}
\end{equation}
Since the linear program has rational coefficients, its optimal solution is guaranteed to be rational. In particular, the quantities $p_k$ and $\overline{q}$ can be chosen as non-negative integers. Consequently,
\begin{eqnarray}
    \sum_{k\in [K]\setminus\overline{\mathcal{S}}}b_k^*= \frac{\sum_{k \in [K]\setminus\overline{\mathcal{S}}} p_k}{\overline{q}} \triangleq \frac{\overline{p}}{\overline{q}}. \label{eq:bk}
\end{eqnarray}
Pick $q$ so that $q > (a^* +b^*)\overline{q} \binom{K \overline{q}}{(a^* + b^*)\overline{q}}$ and operate over $\mathbb{F}_{q}$.
Consider $\overline{p}+(a^*-1)\overline{q}$ i.i.d. uniform variables, $Z_\Sigma={\bf N} = (N_1;\cdots;N_{\overline{p}+(a^*-1)\overline{q}}) \in \mathbb{F}_{q}^{(\overline{p}+(a^*-1)\overline{q}) \times 1}$ and set the key variables as (suppose $\overline{\mathcal{S}} = \{k_1, \cdots, k_{|\overline{\mathcal{S}}|}\}$)
 \begin{eqnarray}
 {Z}_k &=& {\bf F}_k \times {\bf G}_k \times {\bf N}, \forall k \in [K]\setminus\overline{\mathcal{S}} \notag\\
 {Z}_k &=& {\bf H}_k \times {\bf N}, \forall k \in \overline{\mathcal{S}}
 \label{eq:c31} 
 \end{eqnarray}
 where ${\bf F}_k \in \mathbb{F}_{q}^{\overline{q}\times p_k}$, ${\bf G}_k \in \mathbb{F}_{q}^{p_k\times 
 (\overline{p}+(a^*-1)\overline{q})}$, $k \in [K] \setminus \overline{\mathcal{S}}$, ${\bf H}_k \in \mathbb{F}_{q}^{\overline{q}\times 
 (\overline{p}+(a^*-1)\overline{q})}, k \in \overline{\mathcal{S}}$. 
 Without loss of generality, let $\overline{\mathcal{S}} = \{k_1, \ldots, k_{|\overline{\mathcal{S}}|}\}$.
To construct the coefficient vectors, we draw each element of $\{{\bf F}_k,{\bf G}_k\}_{k\in [K]\setminus\overline{\mathcal{S}}},
\{{\bf H}_{k}\}_{k\in \overline{\mathcal{S}}\setminus \{k_1\}}$ independently and uniformly from $\mathbb{F}_{q}$. For correctness, the final vector is defined as the additive inverse of their sum:
 {\setlength{\arraycolsep}{1pt}
 \begin{eqnarray}
{\bf H}_{k_{1}} = - \Bigg( \sum_{k\in[K]\setminus \overline{\mathcal{S}}} {\bf F}_k \times {\bf G}_k + \sum_{k \in \overline{\mathcal{S}} \setminus \{k_{1}\} } {\bf H}_k \Bigg). \label{eq:c32}
 \end{eqnarray}}
By applying a similar argument using the Schwartz-Zippel lemma, we can assert that there exists a specific instantiation\footnote{Observe that, overall, for all matrices ${\bf F}_k^1 \times {\bf G}_k$ and ${\bf H}_k$, the total number of row vectors is $\sum_{k \in [K]\setminus \overline{\mathcal{S}}} p_k + a^* \overline{q} \overset{(\ref{eq:bk})}{=} \sum_{k \in [K]\setminus \overline{\mathcal{S}}} b_k^* \overline{q} + a^* \overline{q} \overset{(\ref{eq:lp})}{=} (a^* + b^* + 1)\overline{q}$. Due to the relation in (\ref{eq:c32}) and the independent selection of ${\bf F}_k, {\bf G}_k, {\bf H}_k$, any subset of $(a^* + b^*)\overline{q}$ rows can be chosen to form a full-rank matrix, ensuring that its determinant polynomial is nonzero.}
of the matrices ${\bf H}_k, {\bf F}_k, {\bf G}_k$ such that (here, ${\bf F}_k^1$ denotes the first $p_k$ rows of ${\bf F}_k$)
\begin{align}
&\text{the row of } {\bf F}_k^1 \times {\bf G}_k \text{ and } {\bf H}_k \text{ are linearly independent,} \notag\\
&\text{if their total rows do not exceed } (a^*+b^*)\overline{q}. \label{eq:sz3}
\end{align}
Moreover,
\begin{eqnarray}
 \sum_{k\in[K]\setminus \overline{\mathcal{S}}} {\bf F}_k \times {\bf G}_k + \sum_{k \in \overline{\mathcal{S}} } {\bf H}_k = {\bf 0}~\Rightarrow~
 \sum_{k\in[K]} Z_k \overset{(\ref{eq:c31})}{=} 0. \label{eq:c33}
\end{eqnarray}
Finally, set $L= \overline{q}$, i.e., ${W_k}=(W_{k,1}; \cdots; W_{k,\overline{q}}) \in \mathbb{F}_{q}^{\overline{q} \times 1}$ and the sent messages as
 \begin{eqnarray}
     {X}_k = {W}_k + {Z}_k, \forall k \in [K].\label{eq:m3}
 \end{eqnarray}
The achieved communication rate is $R_X = L_X / L =\overline{q}/\overline{q}= 1$, and the source key rate is $R_{Z_\Sigma} = L_{Z_\Sigma}/L = (\overline{p} +(a^*-1)\overline{q})/\overline{q} \overset{(\ref{eq:bk})}{=} a^* + \sum_{k\in[K] \setminus \overline{\mathcal{S}}} b_k^*-1 \overset{(\ref{eq:lp})}{=} a^* + b^*$, where the last step is based on a crucial property of the linear program (\ref{lpcoverser1})(\ref{lpcoverser2})(\ref{lpcoverser}), proved next. 
 For correctness, consider any user $u \in [K]$. We have
$\sum_{k \in [K]\setminus\{u\}} X_k + W_u + Z_u = \sum_{k \in [K]\setminus\{u\}} X_k \overset{(\ref{eq:c33})}{=} \sum_{k \in [K]} W_k,$
which ensures that each user can recover the desired sum. The security proof is presented in Section \ref{sec:security}. Next we will prove $\sum_{k\in[K] \setminus \overline{\mathcal{S}}} b_k^*-1 =b^*$.
\begin{lemma}\label{lemma:lp}
For the linear program (\ref{lpcoverser1})(\ref{lpcoverser2})(\ref{lpcoverser}), its optimal value $b^*$ and optimal solution $b^*_k$ satisfy
\begin{eqnarray}
b^* = \sum_{k\in [K]\setminus\overline{\mathcal{S}}} b^*_k - 1. \label{eq:lp}
\end{eqnarray}
\end{lemma}

\proof
First, we prove that $b^* \le \sum_{k\in[K]\setminus\overline{\mathcal{S}}} b_k^* - 1$. Since $b^*$ is the optimal value of the linear program, there exist indices $m_1,n_1,u_1$ such that $b^*=\sum_{k\in (\mathcal{T}_{n_1}\cup\{u_1\})\setminus\overline{\mathcal{S}}} b_k^*$ as attained in (\ref{lpcoverser1}). Then
\begin{eqnarray}
b^* &=& \sum_{k\in (\mathcal{T}_{n_1}\cup \{u_1\})\setminus\overline{\mathcal{S}}} b^*_{k} \label{eq:s3}\\
&=& \sum_{k\in [K]\setminus\overline{\mathcal{S}}} b^*_{k} - \sum_{k\in [K]\setminus (\mathcal{S}_{m_1} \cup \mathcal{T}_{n_1}\cup \{u_1\}) } b^*_{k} \label{eq:s31}\\
&\overset{(\ref{lpcoverser2})}{\leq}& \sum_{k\in [K]\setminus\overline{\mathcal{S}}} b^*_{k} - 1. \label{eq:s32}
\end{eqnarray}
where (\ref{eq:s31}) follows from the identity $\mathcal{T}_{n_1}\cup\{u_1\}=([K]\setminus\overline{\mathcal{S}})\setminus([K]\setminus(\mathcal{S}_{m_1}\cup\mathcal{T}_{n_1}\cup\{u_1\}))$, and (\ref{eq:s32}) holds because for all $m,n,u$, the inequality $\sum_{k\in [K]\setminus(\mathcal{S}_{m}\cup\mathcal{T}_{n}\cup\{u\})} b_k^{*}\ge 1$ is ensured by the feasibility condition in (\ref{lpcoverser2}).

Second, we prove $b^* \geq \sum_{k\in[K]\setminus\overline{\mathcal{S}}} b^*_k - 1$. For the linear program, at least one of constraints (\ref{lpcoverser2}) must be tight (otherwise the binding constraints on $b_k^*$ are all from (\ref{lpcoverser}), i.e., $b^*_k =0$, which violate (\ref{lpcoverser2})). 
Suppose it is for $m_2, n_2,u_2$, such that $\sum_{k\in [K]\setminus(\mathcal{S}_{m_2}\cup\mathcal{T}_{n_2}\cup\{u_2\})} b_k^{*}= 1$, we have
\begin{eqnarray}
1 &=& \sum_{k\in [K]\setminus (\mathcal{S}_{m_2} \cup \mathcal{T}_{n_2}\cup \{u_2\})} b_{k}^* \\
&=& \sum_{k\in [K]\setminus\overline{\mathcal{S}}} b^*_{k} -  \sum_{k\in (\mathcal{T}_{n_2}\cup \{u_2\}) \setminus\overline{\mathcal{S}}} b^*_{k} \label{eq:s41}\\
&\overset{(\ref{lpcoverser1})}{\geq}& \sum_{k\in [K]\setminus\overline{\mathcal{S}}} b^*_{k} - b^*.\label{eq:s42}
\end{eqnarray}
where (\ref{eq:s41}) follows from the identity $[K]\setminus(\mathcal{S}_{m_2}\cup\mathcal{T}_{n_2}\cup\{u_2\}) = ([K]\setminus\overline{\mathcal{S}})\setminus\big((\mathcal{T}_{n_1}\cup\{u_1\})\setminus\overline{\mathcal{S}}\big)$, and (\ref{eq:s42}) holds because for all $m,n,u$, the inequality $\sum_{k\in (\mathcal{T}_{n}\cup\{u\})\setminus\overline{\mathcal{S}}} b_k^* \le b^*$ is guaranteed by the feasibility condition in (\ref{lpcoverser1}), where $b^*$ is the maximum value of $\sum_{k\in(\mathcal{T}_{n}\cup\{u\})\setminus\overline{\mathcal{S}}} b_k^*$ over all $m,n,u$.

The proof of Lemma \ref{lemma:lp} is now complete.

 \subsection{Proof of Security}\label{sec:security} 
Consider any sets $\mathcal{S}_m, \mathcal{T}_n$, and user $\{u\}$ with $m\in[M], n\in[N], u\in[K]$, such that
$|\mathcal{S}_m\cup\mathcal{T}_n\cup\{u\}| \le K-1$. We next show that the security constraint (\ref{security}) is satisfied for all three cases discussed above.
\begin{align}
& I\Big(\left\{W_k\right\}_{k \in \mathcal{S}_m}; \left\{X_k\right\}_{k \in [K]\setminus \{u\}} \Big|\notag\\
&\sum_{k \in [K]} W_k, \left\{W_k, Z_k\right\}_{k \in \mathcal{T}_n},Z_u,W_u \Big)\notag~~\\
\overset{\substack{(\ref{eq:m1})\\(\ref{eq:m2})\\(\ref{eq:m3})}}{=}
& H\Big(\left\{W_k+Z_k\right\}_{k \in [K]} \Big| \sum_{k \in [K]} W_k, \left\{W_k, Z_k\right\}_{k \in \mathcal{T}_n\cup \{u\}} \Big)\notag\\
&-~H\Big(\left\{W_k+Z_k\right\}_{k \in [K]\setminus \{u\}} \Big| \sum_{k \in [K]} W_k, \notag\\
&~~~~~~~\vphantom{ \Big| \sum_{k \in [K]} W_k,}
\left\{W_k, Z_k\right\}_{k \in \mathcal{T}_n\cup \{u\}},\left\{W_k\right\}_{k \in \mathcal{S}_m} \big)\label{eq_thm1_a4}\\
=& H\Big(\left\{W_k+Z_k\right\}_{k \in [K]} \Big| \sum_{k \in [K]} W_k, \left\{W_k, Z_k\right\}_{k \in \mathcal{T}_n\cup \{u\}} \Big)\notag\\
&-~H\Big(\left\{W_k+Z_k\right\}_{k \in \mathcal{S}_m\setminus(\mathcal{T}_n\cup\{u\})} \Big|  \sum_{k \in [K]} W_k,\notag\\
&~~~~
\left\{W_k\right\}_{k \in (\mathcal{S}_m\cup\mathcal{T}_n\cup\{u\})},\left\{Z_k\right\}_{k \in \mathcal{T}_n\cup\{u\}}  \big)\notag\\
&-~H\Big(\left\{W_k+Z_k\right\}_{k \in [K]\setminus(\mathcal{S}_m\cup\mathcal{T}_n\cup\{u\})} \Big| \sum_{k \in [K]} W_k, 
\notag\\
&~~~~
\left\{W_k\right\}_{k \in (\mathcal{S}_m\cup\mathcal{T}_n\cup\{u\})},\left\{Z_k\right\}_{k \in (\mathcal{S}_m\cup\mathcal{T}_n\cup\{u\})}\big)\\
\overset{(\ref{ind})}{=}& H\Big(\left\{W_k+Z_k\right\}_{k \in [K]} \Big| \sum_{k \in [K]} W_k, \left\{W_k, Z_k\right\}_{k \in \mathcal{T}_n\cup \{u\}} \Big)\notag\\
&-H\big(\left\{Z_k\right\}_{k \in \mathcal{S}_m\setminus(\mathcal{T}_n\cup\{u\})} \big|\left\{Z_k\right\}_{k \in \mathcal{T}_n\cup\{u\}}  \big) \notag\\
&-~H\Big(\left\{W_k+Z_k\right\}_{k \in [K]\setminus(\mathcal{S}_m\cup\mathcal{T}_n\cup\{u\})} \Big| \sum_{k \in [K]} W_k, 
\notag\\
&~~~~
\left\{W_k\right\}_{k \in (\mathcal{S}_m\cup\mathcal{T}_n\cup\{u\})},\left\{Z_k\right\}_{k \in (\mathcal{S}_m\cup\mathcal{T}_n\cup\{u\})}\big)\label{security_eq1}\\
\leq&|\mathcal{S}_m\setminus(\mathcal{T}_n\cup\{u\})|L-|\mathcal{S}_m\setminus(\mathcal{T}_n\cup\{u\})|L\\
=&0.
\end{align}
In (\ref{security_eq1}), the difference between the first and the third terms is at most $|\mathcal{S}_m\setminus(\mathcal{T}_n\cup\{u\})|L$, as shown below. The second term also equals $|\mathcal{S}_m\setminus(\mathcal{T}_n\cup\{u\})|L$, which is established in Lemma \ref{lemma:independent}.
\begin{align}
    & H\Big(\left\{W_k+Z_k\right\}_{k \in [K]} \Big| \sum_{k \in [K]} W_k, \left\{W_k, Z_k\right\}_{k \in \mathcal{T}_n\cup \{u\}} \Big)\notag\\
&-~H\Big(\left\{W_k+Z_k\right\}_{k \in [K]\setminus(\mathcal{S}_m\cup\mathcal{T}_n\cup\{u\})} \Big| \sum_{k \in [K]} W_k, 
\notag\\
&~~~~
\left\{W_k\right\}_{k \in (\mathcal{S}_m\cup\mathcal{T}_n\cup\{u\})},\left\{Z_k\right\}_{k \in (\mathcal{S}_m\cup\mathcal{T}_n\cup\{u\})}\big)\\
=& H\big(\left\{W_k+Z_k\right\}_{k \in (\mathcal{S}_m\cup\mathcal{T}_n\cup \{u\})} \big|\notag\\
&~~~\sum_{k \in [K]} W_k, \left\{W_k, Z_k\right\}_{k \in \mathcal{T}_n\cup\{u\}} \Big)\notag\\
    &+~H\Big(\left\{W_k+Z_k\right\}_{k \in [K]\setminus(\mathcal{S}_m\cup\mathcal{T}_n\cup\{u\})} \Big|\sum_{k \in [K]} W_k,\notag\\
    &~~
    \left\{W_k, Z_k\right\}_{k \in \mathcal{T}_n\cup\{u\}},\left\{W_k+Z_k\right\}_{k \in (\mathcal{S}_m\cup\mathcal{T}_n\cup\{u\})} \big)\notag\\
&-~H\Big(\left\{W_k+Z_k\right\}_{k \in [K]\setminus(\mathcal{S}_m\cup\mathcal{T}_n\cup\{u\})} \Big| \sum_{k \in [K]} W_k, 
\notag\\
&~~~~
\left\{W_k\right\}_{k \in (\mathcal{S}_m\cup\mathcal{T}_n\cup\{u\})},\left\{Z_k\right\}_{k \in (\mathcal{S}_m\cup\mathcal{T}_n\cup\{u\})}\big)\\
     \overset{}{\leq}& H\big(\left\{W_k+Z_k\right\}_{k \in (\mathcal{S}_m\setminus(\mathcal{T}_n\cup\{u\}))} \big)\notag\\
    &+~H\Big(\left\{W_k+Z_k\right\}_{k \in [K]\setminus(\mathcal{S}_m\cup\mathcal{T}_n\cup\{u\})} \Big|
       \notag\\
&~~~~~~~~
    \sum_{k \in [K]\setminus(\mathcal{S}_m\cup\mathcal{T}_n\cup\{u\})} \{W_k+Z_k\}\Big)\notag\\
    &-~H\big(\left\{W_k\right\}_{k \in [K]\setminus(\mathcal{S}_m\cup\mathcal{T}_n\cup \{u\})} \big|\left\{Z_k\right\}_{k \in [K]},  
        \notag\\
&~
    \sum_{k \in [K]\setminus(\mathcal{S}_m\cup\mathcal{T}_n\cup \{u\})} W_k,\left\{W_k\right\}_{k \in (\mathcal{S}_m\cup\mathcal{T}_n\cup \{u\})} \Big) \label{eq:ss2}\\
    \overset{(\ref{ind})(\ref{h2})}{\leq}& |\mathcal{S}_m\setminus(\mathcal{T}_n\cup \{u\})|L \notag\\
    &+ \left(\big|[K]\setminus(\mathcal{S}_m\cup\mathcal{T}_n\cup \{u\})\big| - 1\right)L \notag\\
    &- \left(\big|[K]\setminus(\mathcal{S}_m\cup\mathcal{T}_n\cup \{u\})\big| - 1\right)L \label{eq:ss3}\\
    =& |\mathcal{S}_m\setminus(\mathcal{T}_n\cup \{u\})|L.
\end{align}
To obtain the second and third terms in (\ref{eq:ss2}), we use the zero-sum property of the key variables, $\sum_{k\in [K]} Z_k = 0$ (see (\ref{eq:c13}), (\ref{eq:c23}), (\ref{eq:c33})), together with the fact that conditioning cannot increase entropy. In (\ref{eq:ss3}), the first two terms are bounded by counting the corresponding set sizes, while the third term follows from the independence (and uniformity) of $\{W_k\}_{k\in [K]}$ and $\{Z_k\}_{k\in [K]}$.

To complete the proof, we are left to prove the following lemma.
\begin{lemma}\label{lemma:independent}
For all cases of the achievable scheme presented in Section \ref{sec:ach1}, \ref{sec:ach2}, and \ref{sec:ach3}, we have
\begin{align}
&H\left(\{Z_k\}_{k \in \mathcal{S}_m\setminus(\mathcal{T}_n\cup\{u\})} \big|\{Z_k\}_{k \in \mathcal{T}_n\cup\{u\}}  \right)\notag\\
=& |\mathcal{S}_m\setminus(\mathcal{T}_n\cup\{u\})|L, 
\forall m \in [M], n \in [N], u\in [K]. \label{eq:independent}
\end{align}
\end{lemma}

\proof We prove each case one by one.

\noindent\textbf{Subcase 1 of the `Else if' Case (Section \ref{sec:ach1}):}
\begin{align}
&H\left((Z_k)_{k \in \mathcal{S}_m\setminus(\mathcal{T}_n\cup\{u\})} |(Z_k)_{k \in \mathcal{T}_n\cup\{u\}}  \right) \notag\\
=& H\big((Z_k)_{k \in \mathcal{S}_m \cup \mathcal{T}_n\cup\{u\}}\big) - H\left((Z_k)_{k \in \mathcal{T}_n\cup\{u\}}  \right) \\
\overset{(\ref{eq:c11})}{=}& H\big((Z_k)_{k \in (\mathcal{S}_m \cup \mathcal{T}_n\cup\{u\}) \cap \overline{\mathcal{S}} } \big) - H\big((Z_k)_{k \in (\mathcal{T}_n\cup\{u\}) \cap \overline{\mathcal{S}} }  \big) \label{eq:st1}\\
\overset{(\ref{eq:sz1})}{=}&  \big|(\mathcal{S}_m \cup \mathcal{T}_n\cup\{u\}) \cap \overline{\mathcal{S}}\big| L - \big| (\mathcal{T}_n\cup\{u\}) \cap \overline{\mathcal{S}} \big| L \label{eq:st3} \\
=& \big|(\mathcal{S}_m\setminus(\mathcal{T}_n\cup\{u\})) \cap \overline{\mathcal{S}} \big| L= |\mathcal{S}_m\setminus(\mathcal{T}_n\cup\{u\})|L, \notag
\end{align}
where (\ref{eq:st1}) follows from the fact that $Z_k = 0$ for all $k \in [K]\setminus\overline{\mathcal{S}}$ (see (\ref{eq:c11})). To derive (\ref{eq:st3}), we use that $|\mathcal{A}_{m,n,u}| = \big|(\mathcal{S}_m \cup \mathcal{T}_n \cup \{u\}) \cap \overline{\mathcal{S}}\big| \le a^*$ as required in Section~\ref{sec:ach1}, which ensures that the vectors ${\bf H}_k$ involved in the aggregation are linearly independent (see (\ref{eq:sz1})). The last step follows from the fact that $\mathcal{S}_m \subset \overline{\mathcal{S}}$.

\noindent\textbf{Subcase 2 of the `Else if' Case (Section \ref{sec:ach2}):}

We have two sub-cases. 
For the first sub-case, $u' \in \mathcal{T}_n\cup\{u\}$. Then $\big|(\mathcal{S}_m \cup \mathcal{T}_n\cup\{u\}) \cap \overline{\mathcal{S}}\big| < \big|\overline{\mathcal{S}}\big|= a^*$ because otherwise $\big|(\mathcal{S}_m \cup \mathcal{T}_n\cup\{u\}) \cap \overline{\mathcal{S}}\big| = \big|\overline{\mathcal{S}}\big| = a^*$ and according to Definition \ref{def:uni}, we have $u' \in (\mathcal{S}_m \cup \mathcal{T}_n\cup\{u\}) \subset \mathcal{Q}$, which violates our choice of $u'$ to be outside $\mathcal{Q}$ in Section \ref{sec:ach2}. 


We consider two sub-cases. For the first sub-case, $u' \in \mathcal{T}_n \cup \{u\}$. In this case, we must have $\big|(\mathcal{S}_m \cup \mathcal{T}_n \cup \{u\}) \cap \overline{\mathcal{S}}\big| < |\overline{\mathcal{S}}| = a^*$. Otherwise, if $\big|(\mathcal{S}_m \cup \mathcal{T}_n \cup \{u\}) \cap \overline{\mathcal{S}}\big| = |\overline{\mathcal{S}}| = a^*$, then by Definition~\ref{def:uni} it follows that $u' \in (\mathcal{S}_m \cup \mathcal{T}_n \cup \{u\}) \subset \mathcal{Q}$, which contradicts our choice of $u'$ to lie outside $\mathcal{Q}$ in Section~\ref{sec:ach2}.
\begin{align}
& H\left((Z_k)_{k \in \mathcal{S}_m\setminus(\mathcal{T}_n\cup\{u\})} |(Z_k)_{k \in \mathcal{T}_n\cup\{u\}}  \right) 
\notag\\
\overset{(\ref{eq:c21})}{=}& H\big((Z_k)_{k \in (\mathcal{S}_m \cup \mathcal{T}_n\cup\{u\}) \cap (\overline{\mathcal{S}} \cup \{u'\})} \big) 
\notag\\&
~- H\big((Z_k)_{k \in (\mathcal{T}_n\cup\{u\}) \cap (\overline{\mathcal{S}} \cup \{u'\}) }  \big) \label{eq:st11} \\
\overset{(\ref{eq:sz2})}{=}&  \big|(\mathcal{S}_m \cup \mathcal{T}_n\cup\{u\}) \cap (\overline{\mathcal{S}} \cup \{u'\}) \big| L \notag\\
&- \big| (\mathcal{T}_n\cup\{u\}) \cap ( \overline{\mathcal{S}} \cap \{u'\}) \big| L \label{eq:st5} \\
=& \big|(\mathcal{S}_m\setminus(\mathcal{T}_n\cup\{u\})) \cap (\overline{\mathcal{S}} \cup \{u\})\big| L\\
= &|\mathcal{S}_m\setminus(\mathcal{T}_n\cup\{u\})|L,
\end{align}
where (\ref{eq:st11}) follows from the fact that $Z_k = 0$ for all $k \in [K]\setminus(\overline{\mathcal{S}} \cup \{u'\})$ (see (\ref{eq:c21})). Since $\big|(\mathcal{S}_m \cup \mathcal{T}_n \cup \{u\}) \cap \overline{\mathcal{S}}\big| < a^*$, we have $\big|(\mathcal{S}_m \cup \mathcal{T}_n \cup \{u\}) \cap (\overline{\mathcal{S}} \cup \{u'\})\big| \le a^*$. Hence, the first term of (\ref{eq:st5}) is no greater than $a^*$, which allows us to apply (\ref{eq:sz2}) to reduce the rank to the corresponding set cardinality. The last step uses the fact that $u' \in \mathcal{T}_n \cup \{u\}$. For the second sub-case, $u' \notin \mathcal{T}_n \cup \{u\}$ (recall that $u' \notin \overline{\mathcal{S}}$ and $\mathcal{S}_m \subset \overline{\mathcal{S}}$).
{\setlength{\arraycolsep}{1pt}
\begin{align}
&H\left((Z_k)_{k \in \mathcal{S}_m\setminus(\mathcal{T}_n\cup\{u\})} |(Z_k)_{k \in \mathcal{T}_n\cup\{u\}}  \right)\notag\\
\overset{(\ref{eq:c21})}{=}& H\big((Z_k)_{k \in (\mathcal{S}_m \cup \mathcal{T}_n\cup\{u\}) \cap (\overline{\mathcal{S}} \cup \{u'\})} \big) 
\notag\\
&~- H\big((Z_k)_{k \in (\mathcal{T}_n\cup\{u\}) \cap (\overline{\mathcal{S}} \cup \{u'\}) }  \big) \\
\overset{}{=}& H\big((Z_k)_{k \in (\mathcal{S}_m \cup \mathcal{T}_n\cup\{u\}) \cap \overline{\mathcal{S}} } \big) \notag\\
&- H\big((Z_k)_{k \in (\mathcal{T}_n\cup\{u\}) \cap \overline{\mathcal{S}} }  \big) \label{eq:st4}\\
\overset{(\ref{eq:sz2})}{=}&  \big|(\mathcal{S}_m \cup \mathcal{T}_n\cup\{u\}) \cap \overline{\mathcal{S}}\big| L - \big| (\mathcal{T}_n\cup\{u\}) \cap \overline{\mathcal{S}} \big| L
\notag\\
=& |\mathcal{S}_m\setminus(\mathcal{T}_n\cup\{u\})|L,
\end{align}}
where (\ref{eq:st4}) uses the fact that $u' \notin \mathcal{T}_n\cup\{u\}$.

\noindent\textbf{`Otherwise' Case (Section \ref{sec:ach3}):}
\begin{align}
&H\left((Z_k)_{k \in \mathcal{S}_m\setminus(\mathcal{T}_n\cup\{u\})} \big|(Z_k)_{k \in \mathcal{T}_n\cup\{u\}}  \right) \notag\\
=& H\big((Z_k)_{k \in \mathcal{S}_m\cup\mathcal{T}_n\cup\{u\}} \big)
- H\big( (Z_k)_{k \in \mathcal{T}_n\cup\{u\}}  \big)\\
=& H\big((Z_k)_{k \in \left( (\mathcal{S}_m\cup\mathcal{T}_n\cup\{u\}) \cap \overline{\mathcal{S}}\right) \cup \left( (\mathcal{S}_m\cup\mathcal{T}_n\cup\{u\}) \setminus \overline{\mathcal{S}}\right) } \big) 
\notag\\&
- H\big( (Z_k)_{k \in ((\mathcal{T}_n\cup\{u\}) \cap \overline{\mathcal{S}}) \cup ((\mathcal{T}_n\cup\{u\}) \setminus \overline{\mathcal{S}})}  \big) \\
\overset{(\ref{eq:sz3})}{=}& \Big(\big|(\mathcal{S}_m\cup\mathcal{T}_n\cup\{u\}) \cap \overline{\mathcal{S}}\big| \overline{q} + \sum_{k \in (\mathcal{T}_n\cup\{u\}) \setminus \overline{\mathcal{S}}} p_k \Big)  
\notag\\&
- \Big( \big|(\mathcal{T}_n\cup\{u\})  \cap \overline{\mathcal{S}}\big| \overline{q} + \sum_{k \in (\mathcal{T}_n\cup\{u\}) \setminus \overline{\mathcal{S}}} p_k \Big)  \label{eq:sf1} \\
=& |\mathcal{S}_m\setminus(\mathcal{T}_n\cup\{u\})| \overline{q} = |\mathcal{S}_m\setminus(\mathcal{T}_n\cup\{u\})| L.
\end{align}
Using the fact that $L=\overline{q}$ for the scheme in Section~\ref{sec:ach3}, the last step follows. To apply (\ref{eq:sz3}) and obtain (\ref{eq:sf1}), we note that $H((Z_k)_{k\in(\mathcal{T}_n\cup\{u\})\setminus\overline{\mathcal{S}}})$ is determined by the first $p_k$ rows because ${\bf F}_k \times {\bf G}_k$ has rank $p_k$ (see (\ref{eq:c31})). It then suffices to show that the first term is at most $(a^*+b^*)\overline{q}$, and the proof is divided into two sub cases. For the first sub case, when $|(\mathcal{S}_m\cup\mathcal{T}_n\cup\{u\})\cap\overline{\mathcal{S}}|=a^*$, we have $|(\mathcal{S}_m\cup\mathcal{T}_n\cup\{u\})\cap\overline{\mathcal{S}}|\overline{q}+\sum_{k\in(\mathcal{T}_n\cup\{u\})\setminus\overline{\mathcal{S}}}p_k=a^*\overline{q}+\overline{q}\sum_{k\in(\mathcal{T}_n\cup\{u\})\setminus\overline{\mathcal{S}}}b_k^*\le (a^*+b^*)\overline{q}$, where the inequality follows from the maximization objective of the linear program (\ref{lpcoverser1}). For the second sub case, when $|(\mathcal{S}_m\cup\mathcal{T}_n\cup\{u\})\cap\overline{\mathcal{S}}|<a^*$, we obtain $|(\mathcal{S}_m\cup\mathcal{T}_n\cup\{u\})\cap\overline{\mathcal{S}}|\overline{q}+\sum_{k\in(\mathcal{T}_n\cup\{u\})\setminus\overline{\mathcal{S}}}p_k\le (a^*-1)\overline{q}+\overline{q}\sum_{k\in[K]\setminus\overline{\mathcal{S}}}b_k^*\overset{(\ref{eq:lp})}{=}(a^*+b^*-1)\overline{q}$, which is strictly smaller than $(a^*+b^*)\overline{q}$, completing the argument.

\section{Discussion}
This work establishes the exact minimum key size for decentralized secure aggregation with arbitrary collusion and heterogeneous security constraints and shows that the optimal randomness always splits into an integral part and a fractional part determined by a linear program. The fractional component emerges naturally from arbitrary, non-nested security and collusion patterns, reflecting the underlying combinatorial structure. Our achievable scheme matches the converse precisely, proving that the optimal region is both tight and attainable in practice. These results clarify how heterogeneous privacy requirements influence key allocation and when partial security yields meaningful key savings. They also provide actionable guidance for designing efficient decentralized learning systems and motivate extensions to dynamic models and more general network topologies.
\let\url\nolinkurl

\bibliography{references_secagg}
\end{document}